\documentclass[twocolumn,amsmath,amssymb]{revtex4}
\pdfoutput=1

\usepackage{graphicx}
\usepackage{here}
\usepackage{bm}

\newcommand{\bef}{\beta_{\rm f}}
\newcommand{\e}{\epsilon}

\def\cal#1{\mathcal{#1}}
\def\eq#1{Equation~(\ref{#1})}

\def\beq{\begin{equation}}
\def\eeq{\end{equation}}
\def\bea{\begin{eqnarray}}
\def\eea{\end{eqnarray}}

\def\rh{\hat{r}}
\def\kt{k_{\rm B} T}
\def\kb{k_{\rm B}}

\begin{document}

\title{Avoiding unphysical kinetic traps in Monte Carlo simulations of strongly attractive particles}

\author{Stephen Whitelam} 
\author{Phillip L. Geissler}
\affiliation{Department of Chemistry, University of California at Berkeley, and Physical Biosciences and Materials Sciences Divisions, Lawrence Berkeley National Laboratory, Berkeley, CA 94720}

\date{\today}

\begin{abstract}
We introduce a `virtual-move' Monte Carlo (VMMC) algorithm for systems of pairwise-interacting particles. This algorithm facilitates the simulation of particles possessing attractions of short range and arbitrary strength and geometry, an important realization being self-assembling particles endowed with strong, short-ranged and angularly specific (`patchy') attractions. Standard Monte Carlo techniques employ sequential updates of particles and can suffer from low acceptance rates when attractions are strong.  In this event, collective motion can be strongly suppressed. Our algorithm avoids this problem by proposing simultaneous moves of collections (clusters) of particles according to gradients of interaction energies. One particle first executes a `virtual' trial move. We determine which of its neighbours move in a similar fashion by calculating individual bond energies before and after the proposed move. We iterate this procedure and update simultaneously the positions of all affected particles. Particles move according to an approximation of realistic dynamics without requiring the explicit computation of forces, and without the step size restrictions required when integrating equations of motion. We employ a size- and shape-dependent damping of cluster movements, motivated by collective hydrodynamic effects neglected in simple implementations of Brownian dynamics. We discuss the virtual-move algorithm in the context of other Monte Carlo cluster-move schemes, and demonstrate its utility by applying it to a model of biological self-assembly.
\end{abstract}

\pacs{02.70.Tt,87.53.Wz}
\maketitle

\section{Introduction}
A standard Monte Carlo simulation of a system of particles consists of a sequence of moves of  individual particles. If these moves are proposed and accepted according to the principle of detailed balance (or balance~\cite{deem,young} or superdetailed balance~\cite{super_duper}), the system will eventually relax to thermal equilibrium~\cite{Daan_book}. For some models, this approach even provides an approximation to the dynamics that the corresponding physical system would execute~\cite{mc_dynamics1,mc_dynamics2,ludo}. However, such approximations break down when relaxation requires the movement of particles in concert, particularly in the presence of very strong interactions. This is the case, for example, when small ions of opposite charge bind together tightly~\cite{ions}.

\begin{figure}
\includegraphics[width=5.5cm]{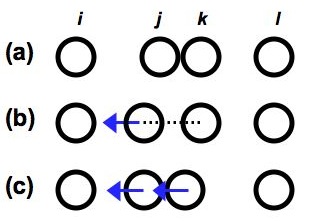} 
\caption{\label{figzero} An illustration of the difficulty encountered by standard Monte Carlo methods in the presence of very strong, short-ranged interactions. Here only particles $j$ and $k$ interact, and do so strongly. Single-particle Monte Carlo schemes generate a diffusion of the dimer $jk$ via relative moves of $j$ and $k$. Because such moves (e.g. {\bf a} $ \to $ {\bf b}) are suppressed by the exponential of the resulting energy change, collective modes of motion are under-represented for any acceptance rate less than unity. Moving particles $j$ and $k$ simultaneously ({\bf a} $\to$ {\bf c}) can restore collective motion. We shall discuss how this can be done in order to approximate a realistic dynamics.}
\end{figure}

In such situations, the suppression of collective modes of motion exhibited by standard Monte Carlo algorithms affects both intra-cluster relaxation and whole-cluster diffusion. As an example of the latter we show in Figure~\ref{figzero} four particles equipped with strong pairwise interactions. Only two particles ($j$ and $k$) are close enough to interact. An `overdamped' (Langevin) dynamics would see particle positions evolve according to both deterministic forces, derived from pairwise interactions, and random buffeting forces designed to model solvent fluctuations. The random forces induce collective diffusion of isolated clusters (such as the dimer $jk$). Molecular dynamics simulations naturally accommodate collective motion by updating {\em simultaneously} the position of every particle, according to both deterministic and random forces. A standard Monte Carlo simulation, however, employs {\em sequential} updates of individual particle positions. The effects of the deterministic and random buffeting forces are then inextricably linked. Potential energy gradients dictate that moves of one particle relative to another (e.g. the displacement that changes configuration {\bf a} to configuration {\bf b}) are suppressed by the exponential of the change in interaction energy. If this change is but a few multiples of $k_{\rm B }T$, such a move is unlikely to be accepted. Since the random buffeting force is modeled by the random motion of individual particles, rejecting such motion has the undesirable effect of annulling the instantaneous buffeting force experienced by that particle. This kinetic `trap' leads to a suppression of collective diffusion, and hence to an unphysical dynamics.

In a similar manner, internal cluster relaxation involving collective modes of motion is under-represented by making sequential moves of single particles. A realistic description of collective structural rearrangement is necessary to model many phase transitions, such as the microphase separation of colloids at low temperature, and examples of self-assembly, e.g. of large proteins called chaperonins~\cite{chap1a, chap1b, chap1c}; various nanoparticles~\cite{Glotzer0, Glotzer1, Glotzer2}; and virus capsids~\cite{Rapaport,Hagan}.

For sufficiently small displacements, single-particle moves can in principle retain acceptance rates large enough to permit some collective modes of relaxation. However, when interactions are not only strong but also short-ranged, the displacement scale required to ensure acceptable rates of collective motion is invariably so small that configuration space cannot be explored in a reasonable time. 

Compounding these difficulties in modeling a natural dynamics is the fact that even in the limit of unit acceptance rate the diffusion associated with single-particle Monte Carlo protocols represents only a rough approximation of physical motion. Single-particle moves (and indeed standard implementations of Brownian dynamics) induce motion of an isolated cluster with a translational diffusion constant that scales as the reciprocal of the number of constituents of the cluster, and a rotational diffusion constant proportional to the reciprocal of the cluster's moment of inertia~\cite{footnote0}. By contrast, Stokes' law for a body moving through a viscous medium implies translational and rotational diffusion properties depending on the first and third powers, respectively, of the cluster's characteristic radius. 

These limitations can be overcome in principle by proposing simultaneous moves of clusters of particles (e.g. the move that takes state {\bf a} to state {\bf c} in Figure~\ref{figzero}). Existing cluster algorithms~\cite{Panag, Liu1, SW,big_bad_wolff,wu} have been used with great success to simulate many complex systems~\cite{review}. Such algorithms define a cluster to be moved by recursively linking particle pairs with a given probability. For those algorithms that effect local moves of particles, the simultaneous displacement of collections of particles allows the restoration of diffusive modes of motion suppressed by single-particle schemes.

However, such algorithms restore diffusive motion by identifying and moving clusters based on the properties of particles (energy or proximity) in their {\em current} configuration, without regard for the changes induced by the proposed move. As a consequence, these algorithms tend to under-represent internal collective motion: particles interacting strongly (or particles in close proximity) will be displaced collectively, and will therefore not move relative to each other. This leads to the development of severe kinetic traps. In Appendix A we discuss how generalizing these algorithms to permit the identification of clusters with greater flexibility can lead to efficent collective re-arrangement. However, while equilibration can in this manner be facilitated, identifying and moving clusters according to properties of the current configuration does not restore a physical {\em dynamics}: particles do not explore local gradients of potential energy in a realistic manner. We discuss this situation in more detail in Appendices A and B. 

To permit both realistic diffusive motion and collective internal re-arrangement we propose defining and moving clusters on the basis of {\em gradients} in potential energy. In this paper we introduce a `virtual-move' Monte Carlo (VMMC) scheme designed to approximate a realistic dynamics for particles possessing strong, short-ranged interactions. In our scheme, one particle first executes a `virtual' trial move. We determine which of its neighbours move in a similar fashion by calculating bond energies before and after the proposed move. We iterate this procedure from all `recruited' neighbours and stop when no more particles show a tendency to move. We then update simultaneously the positions of all affected particles.

 The VMMC scheme is designed to facilitate the simulation of components that spontaneously self-assemble. Such components typically possess pairwise attractions of arbitrary strength, possibly very short range, and a high degree of angular specificity (`patchiness'). The algorithm executes collective updates of particles in a manner designed to approximate realistic particle motion. It permits a basic particle displacement scale larger than can be used, for example, when integrating equations of motion. Furthermore, cluster moves may be performed so as to respect Stokes' law, offering the possibility of exceeding the dynamical realism of simple implementations of Brownian dynamics. 

This paper is organised as follows. In Section \ref{section2} we show that by proposing collective moves of particles based upon individual bond energy gradients one can evolve in an approximately realistic (and computationally efficient) way a system of strongly pairwise-interacting particles. In Section~\ref{section3} we generalize the scheme to permit the use of distinct real and virtual moves, thereby allowing precise control of relative rotational and translational motion. We show that in this way we can mimic the types of relaxation observed in Brownian dynamics simulations of isotropic Lennard-Jones particles in two dimensions. In Section~\ref{section4} we apply the VMMC algorithm to a system of self-assembling components designed to model the aggregation of protein complexes called chaperonins. We conclude in Section \ref{conc}.

\section{A `virtual-move' Monte Carlo cluster algorithm}
\label{section2}
\subsection{Making collective moves}

In this section we introduce our central result, a virtual-move Monte Carlo algorithm. This algorithm is designed to approximate physical dynamics by moving interacting bodies either individually or in concert according to individual bond energy gradients. This procedure may be regarded as a particle-based adaptation of the Wolff cluster algorithm~\cite{big_bad_wolff}, and may be used on- or off-lattice. It is similar in some respects to the algorithm described in Ref.~\cite{Liu1}, but differs in that here proposed moves are chosen to correspond as closely as possible to realistic particle displacements. 

As discussed in the introduction, making sequential moves of single particles leads to a suppression of collective modes of motion. To correct this deficiency it is necessary to effect moves of particles in concert. We shall explain in this subsection how this can be done in general. In the following subsection we describe the specifics of the virtual-move scheme. In brief, we effect cluster displacements by first proposing a move of a single particle. If the change in energy of interaction between that particle and its neighbours is unfavourable, those neighbours are `recruited', adopting the move of the first particle as their own. We continue this recruitment until no further particles show a tendency to move. Before describing this scheme in detail we shall set up our notation and nomenclature.

\begin{figure}
\includegraphics[width=8cm]{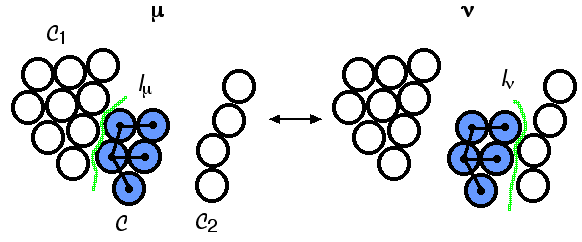} 
\caption{\label{figone} 
An illustrative collective move. We define a pseudocluster ${\cal C}$ (shaded particles) using an iterative linking scheme (see text). A particular realization of links is denoted by bold black lines. Failed links internal to the pseudocluster are not shown. We subject the pseudocluster to a rotation or a translation (or both). Here a proposed translation moves ${\cal C}$ from contact with cluster ${\cal C}_1$ in state $\mu$ to contact with cluster ${\cal C}_2$ in state $\nu$. The interface between the pseudocluster and its environment in state $\alpha$ is labelled $I_{\alpha}$, and is defined by the set of all pairwise interactions between white and shaded particles.  }
\end{figure}

We consider in $d$ dimensions a system of $N$ particles having radius $R_0$ and interacting by way of a short-ranged, pairwise potential $\e_{ij}$. In Figure \ref{figone} we show a typical cluster move (and its reverse) from state $\mu$ to state $\nu$. We define two particles to be contiguous if their centres are separated by a distance less than or equal to $r_c$, the interaction range of the potential. We define a {\em physical cluster} (abbreviated simply to `cluster') as a group of contiguous particles. We shall use an iterative linking procedure, described below, to select a group of particles, $\cal{C}$. This group can range in size from a single particle to any one cluster in the system. We shall call this group of particles a {\em pseudocluster}. In Figure \ref{figone} the pseudocluster ${\cal C}$ is shaded. We perform on ${\cal C}$ either a rigid-body rotation or a translation in an arbitrary direction (or, if desired, a linear combination of a translation and a rotation). In the figure, one such translation happens to bring the pseudocluster into contact with the cluster ${\cal C}_2$. 

To define a pseudocluster ${\cal C}$ we select from the system a `seed' particle $i$. The seed is the first member of the pseudocluster. We shall attempt to `recruit' particles as members of the pseudocluster by forming `links' from members of the pseudocluster to neighbouring particles. We start by forming a link with probability $p_{ij}(\mu \to \nu)$ between $i$ and any particle $j$ with which $i$ interacts. The form of $p_{ij}$ shall be specified later. If this link forms we regard $j$ as a member of the pseudocluster. We continue the linking procedure until links have been proposed exactly once from each member $m$ of the pseudocluster to all particles with which $m$ interacts.

Linking particles in this manner results in the pseudocluster ${\cal C}$ being selected and moved with probability
\beq
\label{factor1}
W_{\rm gen}(\mu \to \nu)= P_{\rm seed}(\mu) P_{\rm displace}(\cal{C};\mu \to \nu).
\eeq
Here $P_{\rm seed}(\mu)$ accounts for the likelihood of choosing a seed particle in the pseudocluster ${\cal C}$. The factor $P_{\rm displace}(\cal{C};\mu \to \nu)$ returns the probability (given a seed particle) of building the pseudocluster in state $\mu$ and moving it such that the proposed new state is $\nu$. It depends on two factors:
\beq
\label{factor2}
P_{\rm displace}(\cal{C};\mu \to \nu)= \sum _{{\cal R}}^{{\cal C}}  \prod_{\mu \to \nu} q_{ij}(\mu \to \nu)\prod^{{\cal R}}_{\langle i j \rangle_{\ell}} p_{ij}( \mu \to \nu).
\eeq
The first factor in Equation~(\ref{factor2}) concerns the link-failure probabilities $q_{ij}(\mu \to \nu) \equiv 1-p_{ij}(\mu \to \nu)$. The product $\prod_{\mu \to \nu}$ runs over all links which must {\em not} form in order to move from state $\mu$ to state $\nu$; unformed links external to the pseudocluster define the interface labeled $I_{\mu}$ in Figure~\ref{figone}. The second factor in (\ref{factor2}) is the probability of forming links (labelled $\ell$) between the constituent monomers of ${\cal C}$. In general, there are many distinct realizations $\cal{R}$ of links (and failed links) leading to the same chosen pseudocluster (denoted by $\sum_{{\cal R}}^{{\cal C}}$), and the product runs over links comprising one such realization. Again in general, the probability of forming links between particles such that the shaded cluster is the chosen pseudocluster depends on whether one is executing the forward or reverse move. 

The concept of detailed balance ensures that a system evolves towards equilibrium by requiring that the rates $W$ for passing between any two states $\mu$ and $\nu$ satisfy
\beq
\label{detbal}
\rho(\mu) W(\mu \to \nu) = \rho(\nu) W(\nu \to \mu).
\eeq
Here $\rho(\alpha) = Z^{-1} e^{-\beta H_{\alpha}}$ is the Boltzmann weight with respect to the system's Hamiltonian in state $\alpha\in \{\mu,\nu\}$, $H_{\alpha}$; $\beta \equiv 1/T$ is the reciprocal temperature (we adopt units such that $\kb = 1$); and $Z$ is the partition function. Calculating the rates $W$ would require the enumeration of all possible ways of linking together monomers in order to generate a given pseudocluster. For large systems this is not feasible. Instead, we use an iterative procedure to identify one particular realization of links and failed links, and compute the likelihood that each of these links (or failed links) will form when making the reverse move. We then impose the requirement
\beq
\label{superdetbal}
\rho(\mu) W(\mu \to \nu|{\cal R}) = \rho(\nu) W(\nu \to \mu| \cal{R}),
\eeq
where $W(\mu \to \nu |{\cal R})$ is the rate for passing from state $\mu$ to state $\nu$, given a realization $\cal{R}$ of links and failed links. This is the condition of superdetailed balance~\cite{super_duper}.
The realization ${\cal R}$ must include the {\em direction} in which links are formed (e.g. from particle $i$ to $j$, or vice versa).
 
We note finally that the rate for passing between states is the product of the rates of proposing (generating) and of accepting such a move: $W= W_{\rm gen} \times W_{\rm acc}$. 

Combining the results of this section fixes the ratio of acceptance rates for forward and reverse moves. We choose as the acceptance rate for the move $\mu \to \nu$
\bea
\label{factor3}
&\,&W_{\rm acc}\left(\mu \to \nu|{\cal R}\right) \nonumber \\
&=& \Theta \left(n_{\rm c} -n_{{\cal C}} \right) \cal{D}(\cal{C}) \, \min \left\{ 1, \frac{P_{\rm seed}(\nu)}{P_{\rm seed}(\mu)} {\rm e}^{-\beta(E_{\nu}-E_{\mu})} \nonumber \right. \\ 
&\times& \left. \frac{\prod_{\nu \to \mu} q_{ij}(\nu \to \mu)}{\prod_{\mu \to \nu} q_{ij}(\mu \to \nu) } \prod^{{\cal R}}_{\langle i j \rangle_{\ell}} \frac{p_{ij}( \nu \to \mu)}{p_{ij}( \mu \to \nu)} \right\} .
\eea
The factors outside the `minimum' function are not required to enforce balance (and therefore to ensure sampling the correct equilibrium distribution), but are chosen to preserve as closely as possible a realistic dynamics. The first factor in Equation~(\ref{factor3}) is used in conjunction with an `early rejection' termination of the link-forming procedure if we recruit more particles than is `physical'. The function $\Theta$ returns zero if the number of particles in the pseudocluster, $n_{{\cal C}}$ is greater than a specified cutoff $n_{\rm c}$, and unity otherwise. In order to propose moves of clusters of arbitrary size with correct frequency it is necessary to suppress moves of clusters of size $n_{{\cal C}}>1$ by drawing the cutoff $n_c$ from a particular distribution. Otherwise, large clusters move in general more frequently than small clusters. It is crucial to enforce this frequency-correction {\em during} the link-forming procedure; we would waste much time by simply rejecting moves of large clusters with high probability. 

The factor $\cal{D}(\cal{C}) \leq 1$ encodes the diffusion properties of the pseudocluster ${\cal C}$. One of the advantages of the Monte Carlo cluster-move framework is that we know in advance the size and shape of the aggregate ${\cal C}$ whose displacement we are considering. We can then enforce a hydrodynamic damping of translations and rotations by rejecting moves of clusters with a factor that depends on the size and shape of the aggregate. In Section~\ref{section3} we discuss how to {\em scale} collective translations and rotations instead of simply rejecting collective moves with a specified probability.

The first factor (of the second argument) within the braces accounts for the likelihood of picking a given particle $i$ as a seed in states $\mu$ and $\nu$. We shall choose the seed particle uniformly from any in the system, giving $P_{\rm seed}(\mu)=P_{\rm seed}(\nu)$.

To derive the final factor of the first line within the braces we have used the result that the internal pseudocluster energies in states $\mu$ and $\nu$ are identical, and so the ratio of Boltzmann weights $\rho(\nu)/\rho(\mu)$ reduces to ${\rm e}^{-\beta \left( E_{\nu}- E_{\mu} \right)}$. Here $E_{\alpha}$ is the interfacial energy between the cluster and its environment in state $\alpha$, namely, the sum in that state of all pairwise energies between white and shaded particles. 

The final line of Equation~(\ref{factor3}) contains the link-failure $(q_{ij})$ and link-forming $(p_{ij})$ factors. These depend on the specific choice of the linking probability $p_{ij}(\mu \to \nu)$. In the following subsection we discuss in detail one such choice. Note that the link-failure products run over failed links both internal to and external to the pseudocluster. The link-making factor is evaluated for a given realization ${\cal R}$ of links.

\subsection{Making collective moves according to potential energy gradients}

With the Monte Carlo cluster-move framework described, we turn to a specific choice for the linking probability $p_{ij}(\mu \to \nu)$. To enforce an approximate dynamical realism this probability must depend on the proposed move connecting the initial state $\mu$ to the final state $\nu$. In Appendices A and B we discuss how linking particles according only to properties of the {\em initial} state leads to a dynamics that is not physical.

Our virtual-move scheme is as follows. We start in state $\mu$. We select uniformly a pseudocluster `seed' particle $i$ having coordinates (position and orientation) $x_i$, and assign to this seed a move {\em map}. This map defines a random translation or rotation about an axis through the centre of the seed (or, if desired, a linear combination of both). We denote by $x_i'$ the coordinates of $i$ following application of this map. We execute a `virtual' move of the seed under the map, and look to those particles $\{j\}$ with which the seed interacts in state $\mu$. We link a given particle $j$ to the seed with a probability $p_{ij}(\mu \to \nu)$ that depends on the energy of the relevant bond before {\em and after} the move of the seed:
\bea
\label{virtual_link}
p_{ij}(\mu \to \nu)&=&\Theta \left(n_{\rm c} -n_{{\cal C}}  \right) \nonumber \\
&\times& \cal{I}_{ij}^{(\mu)} \textnormal{max}\left(0,1-{\rm e}^{\beta E_{\rm c}(i,j)-\beta E_{\rm I}(i,j)}\right).
\eea
The energy 
\beq
\label{virtual_linkb}
E_{\rm c}(i,j) \equiv \epsilon_{ij}\left(x_i',x_j'\right) = \epsilon_{ij}\left(x_i,x_j\right)
\eeq
 is the energy of the bond $ij$ following a {\em collective} virtual move of $i$ and $j$ (where each move according to the same map). Because the map defines either a rigid-body rotation or translation, the particles do not move relative to each other, and so this bond energy is identical to that in the starting state $\mu$. This gives rise to the second equality in Equation~(\ref{virtual_linkb}). The term
 \beq
 E_{\rm I}(i,j) = \epsilon_{ij}\left(x_i',x_j\right)
 \eeq
 is the bond energy following an individual virtual move of $i$, and no move of $j$ (the bond energy in the proposed state $\nu$). 
 
We define the `interaction' term $\cal{I}_{ij}^{(\mu)}$ in Equation~(\ref{virtual_link}) to be unity if in state $\mu$ particles $i$ and $j$ are deemed to be interacting (and therefore eligible for the linking procedure), and zero otherwise. We define interacting particle pairs as those whose centres are separated by less than the range of the interaction. We shall see later that we must account separately, via the overall acceptance rate, for certain particle pairs that start or end a move in a `noninteracting' configuration.

The first factor in Equation~(\ref{virtual_link}) enforces an `early rejection' of the link-forming procedure. Because collective moves can in principle be initiated from any particle, those particles residing in large clusters have a greater chance of changing position than do isolated particles. We account for this by suppressing the rate for moves of clusters of size $n_{{\cal C}}$ by a factor of $1/n_{{\cal C}}$. It is not efficient to engineer this suppression via the acceptance rate. {\em A priori} we do not know how many particles will be assigned to a pseudocluster. We would waste much effort by building large pseudoclusters only to reject their move with high probability. Instead, we suppress the {\em generation} rate for pseudocluster construction. For each move we draw a cutoff $n_{\rm c}$ from the distribution $Q(n_{\rm c})\propto n_{\rm c}^{-1}$, and abort the link formation procedure if the pseudocluster size exceeds $n_{\rm c}$. We then cancel the move, as indicated by the prefactor in \eq{factor3}. In this way we ensure (with reasonable computational efficiency) that particles experience positional updates with approximately equal frequency, a requirement necessary for dynamical fidelity. The early-rejection procedure can be further modified to account for the diffusion properties of clusters (see following section).
 
If a link is formed, the linkee particle $j$ is `recruited' to the pseudocluster, and adopts the move map of the seed. The linker particle $i$ is then returned to its original position. We next perform a `reverse' virtual move of the linker particle and record the probability $p_{ij}(\nu \to \mu)$ that the link $ij$ would form were the linker particle translating in the opposite direction (and/or rotating with the opposite sense). This factor will be used to enforce balance in the manner shown in Equation~(\ref{factor3}); it suppresses, for example, collective moves of hard particles with no attractive interactions~\cite{krauth}.

We continue this scheme hierarchically, attempting to link (exactly once) each member $m$ of the pseudocluster to any unlinked particle with which $m$ interacts. We stop when no further links form. We then update simultaneously the positions of all linked particles, moving the pseudocluster as a rigid body, and evaluate the Monte Carlo acceptance probability.

In intuitive terms, with probability $p_{i j}(\mu \to \nu)$ we propose a move of $i$ and $j$ in concert, according to a move map specifying a change in position of $i$ from a state $\mu$ to a state $\nu$. With the complementary probability we instead propose a move of $i$ relative to $j$. This is the key difference from a standard single-particle MC scheme: here, if the move of particle $i$ relative to $j$ is rejected, $i$ and $j$ are moved in concert. 

\begin{figure}
\includegraphics[width=8cm]{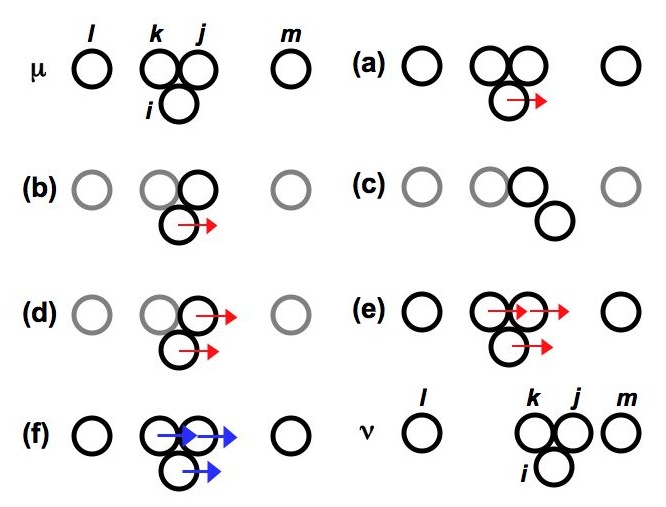} 
\caption{\label{figseven} An illustration of virtual-move Monte Carlo. From starting state $\mathbf{\mu}$ the seed particle $i$ is assigned a virtual move map, denoted by a thin arrow $\mathbf{(a)}$. We propose a link between $i$ and $j$ by calculating the energy difference of the bond $ij$ before and after the virtual move of $i$ $\mathbf{(b,c)}$. In this example a link forms, and $j$ adopts the virtual move map of $i$; the latter is returned to its original position $\mathbf{(d)}$. We iterate this procedure until no more links form $\mathbf{(e)}$, and displace all linked particles simultaneously (bold arrows denote real, not virtual, moves): $\mathbf{(f)}\to \nu$. The new configuration $\nu$ is proposed as the final state, and the Monte Carlo acceptance probability is evaluated.}
\end{figure}

The scheme is best explained using an example, which we show in Figure~\ref{figseven}. We consider five particles, $i$, $j$, $k$, $l$ and $m$, endowed with short-ranged, orientation-dependent pairwise interactions (particle orientations not shown). Bonds $ij$, $ik$ and $jk$ are strong, possessing a large negative energy $\e$, while bond $kl$ is weak, possessing an energy of interaction $\delta \approx 0$. There is no interaction between $m$ and any other particle. 

We first choose a seed particle, say $i$. To $i$ we assign a move map, denoted by a thin arrow ({\bf a}). The map here defines a rightward translation. We shall denote the state of particle $\alpha$ by $x_{\alpha}=\{\mathbf{r}_{\alpha}, \mathbf{S}_{\alpha}\}$ (which enodes the position $\mathbf{r}_{\alpha}$ and orientation $\mathbf{S}_{\alpha}$ of the particle). After executing its virtual move, the state of particle $\alpha$ is $x_{\alpha}'$.

To determine whether $i$ moves individually or in concert with other particles, we form links between $i$ and each particle with a probability given by Equation~(\ref{virtual_link}). In our example the linking procedure unfolds as follows. We propose a link between $i$ and $j$, and so consider only these particles. We calculate the initial energy of the bond $ij$, $E_{\rm c}$, and find it to be large and negative $E_{\rm c}=\e \ll 0$ $\mathbf{(b)}$. We move $i$ under its map and calculate the new energy of the bond $\mathbf{(c)}$. Because attractions are short-ranged we find this energy to be zero. We link $i$ and $j$ with probability $p_{ij}(\mu \to \nu) = 1-\rm{e}^{\beta \e} \approx 1$. In our example this link is accepted, and so $j$ adopts the move map of $i$. The latter is returned to its original position $\mathbf{(d)}$. We then execute a `reverse' virtual move of $i$ (not shown), and record the reverse linking probability $p_{ij}(\nu \to \mu)$. We next propose a link between $j$ and $k$, and find that it is also accepted $\mathbf{(e)}$. Again, we record the reverse linking probability. We continue the link-forming procedure by testing the interaction $kl$; this weak bond is linked with probability $p_{kl}\approx \delta/\kt \approx 0$. In our example this link is not formed. The virtual-move linking procedure is now finished. The final virtual moves are adopted as real moves (thin arrows become bold arrows, panel $\mathbf{(f)})$ and all particles in the chosen pseudocluster ($i,j,k$) are displaced simultaneously, according to their map (defining a new state, $\nu$). We propose this state $\mathbf{\nu}$ as the final state, and evaluate the Monte Carlo acceptance factor. 

\begin{figure}
\includegraphics[width=8cm]{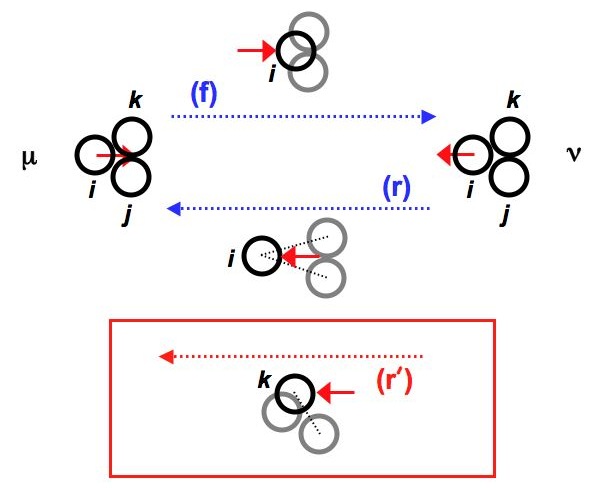} 
\caption{\label{figreverse} Ensuring reversibility for cluster moves. We require that the likelihood of a given particle, say $i$, `pushing' ({\bf f}) and `pulling' ({\bf r}) its host cluster is identical. To do so, every time we link two particles we record the likelihood that the link would have formed had the reverse move been proposed. In the reverse move, the linking particle translates in the opposite direction (and/or rotates with the opposite sense). We account for this probability difference via the overall acceptance rate, ensuring superdetailed balance. The acceptance rate for such collective moves is high if the binding energy of particles is high (the regime of interest), and vanishes in the limit of vanishing interaction strength~\cite{krauth}. We require that the reverse move be initiated by the seed particle of the forward move, because moves having distinct seeds (compare {\bf f} and {\bf r}$'$) cannot in general balance each other.}
\end{figure}

The `reverse' virtual move described above is used to ensure superdetailed balance by suppressing the likelihood of a move $\mu \to \nu$ by the product over all links of terms of the form $p_{ij}(\nu \to \mu)/p_{ij}(\mu \to \nu)$, as shown in Equation~(\ref{factor3}). The necessity of this procedure can be understood in intuitive terms by referring to Figure ~\ref{figreverse}. In this example, the forward move {\bf f} is initiated by displacing the seed $i$ rightwards. This move causes $i$ to overlap particles $j$ and $k$. If these overlaps are `hard' (infinitely unfavourable energetically), links $ij$ and $ik$ form with certainty; we do not need to propose a link between $k$ and $j$. The chosen pseudocluster ${\cal C}$ is the trimer $ijk$, and the proposed final state follows by displacing ${\cal C}$ rightwards.

But now we encounter a problem generating the reverse move. We cannot do so by displacing, for example, $k$ to the left (move {\bf r}${\mathbf '}$). This move will induce a link with $i$ with unit probability, but $j$ will be added to the pseudocluster with a probability depending on its energies of interaction with $i$ and $k$. We shall require that the reverse move be initiated by the seed particle that began the forward move. However, displacing $i$ to the left (`reverse' move {\bf r}) will result in links with $j$ and $k$ only with probability $1-e^{-\beta \Delta \epsilon_{i\alpha}}$, where $\alpha \in \{j,k\}$ and $\Delta \epsilon_{i\alpha}$ is the change in binding energy between $i$ and $\alpha$. To ensure that rightward and leftward moves of ${\cal C}$ occur with equal probability, given that the seed particle is $i$, we must suppress the likelihood of the forward (rightward) move by a factor $(1-e^{-\beta \Delta \epsilon_{ij}}) (1-e^{-\beta \Delta \epsilon_{ik}})$. For strong attractions (the regime in which we are interested) this factor is close to unity; for hard particles with no attractions it is zero, and we must reject the forward move~\cite{krauth}. We thus ensure that the probability of a given particle `pushing' or `pulling' its host cluster is identical~\cite{footnote2}.

The acceptance probability for the virtual-move procedure follows from Equations~(\ref{factor3}) and (\ref{virtual_link}). We first evaluate the link-failure factors $q_{ij}$. Consider a move from state $\mu$ to state $\nu$. The probability of {\em not} linking a particle $j$ (that is not a member of the chosen pseudocluster) to a particle $i$ (a member of the pseudocluster) is, from Equation~(\ref{virtual_link}),
\beq
\label{smithers1}
q_{ij}(\mu \to \nu)= \cal{I}_{ij}^{(\mu)} \min\left(1,e^{\beta \epsilon_{ij}^{(\mu)}-\beta \epsilon_{ij}^{(\nu)}} \right) + 1-\cal{I}_{ij}^{(\mu)}.
\eeq
Here $\epsilon_{ij}^{(\alpha)}$ denotes the pairwise energy of the bond $ij$ in state $\alpha$. Recall that the factor $\cal{I}_{ij}^{(\mu)}$ is unity if in state $\mu$ particles $i$ and $j$ lie close enough to interact, and zero otherwise.

The factor corresponding to (\ref{smithers1}) for the reverse move follows by interchanging the labels $\mu$ and $\nu$. Hence we have for external failed links
\bea
\label{paul_simon}
 \frac{\prod_{\nu \to \mu}^{\rm ext.} q_{ij}(\nu \to \mu) }{\prod_{\mu \to \nu}^{\rm ext.} q_{ij}(\mu \to \nu)} &=& \prod_{\langle i j \rangle'} e^{\beta \epsilon_{ij}^{(\nu)}-\beta \epsilon_{ij}^{(\mu)}},
\eea
where $\langle ij \rangle'$ denotes all shaded-white pairs {\em except} those which fall into two classes. The first class consists of those particle pairs that do not interact in state $\mu$, but which move uphill in energy upon going from state $\mu$ to state $\nu$. Since we define noninteracting particle pairs (via the interaction criterion $\cal{I}_{ij}^{(\mu)}$) as those that in state $\mu$ lie too far apart to possess any energy of interaction, moving uphill in energy means that these pairs end the move (state $\nu$) with positive energy (we shall refer to particles having positive energy of interaction as `overlapping' particles). The second class of particle pairs are those that move downhill in energy upon going from state $\mu$  to state $\nu$, but do not interact in state $\nu$. For the interaction criterion as defined this means those particle pairs that in state $\mu$ overlap each other, but which finish their move (in state $\nu$) outside the interaction region. Note that the second class cannot exist if the potential is composed of a purely attractive piece plus a hard-core repulsion, because then no particle pairs start in an overlapping position.

Our final acceptance rate is then
\bea
\label{accept}
&\,&W_{\rm acc}(\mu \to \nu|\cal{R}) = \nonumber \\
&\,& \Theta \left(n_{\rm c} -n_{{\cal C}} \right) \cal{D}(\cal{C}) \min \left\{1, \prod_{\langle i j \rangle_{{\rm n}\leftrightarrow {\rm o}}} e^{-\beta\left( \epsilon_{ij}^{(\nu)} -\epsilon_{ij}^{(\mu)} \right)}  \right. \nonumber \\ 
&\times& \left. \prod^{{\cal R}}_{\langle i j \rangle_{{\rm f}}} \frac{ q_{ij}( \nu \to \mu)}{q_{ij}( \mu \to \nu)}  \prod^{{\cal R}}_{\langle i j \rangle_{\ell}} \frac{ p_{ij}( \nu \to \mu)}{p_{ij}( \mu \to \nu)} \right\}.
\eea
The link-failure factors external to the pseudocluster have canceled the ratio of Boltzmann bond weights, {\em except} those corresponding to particle-pairs $\langle ij \rangle_{{\rm n} \leftrightarrow {\rm o}}$ falling in either of the two classes defined above: class 1 contains those particle pairs that start ($\mu$) in a noninteracting configuration and end ($\nu$) in an overlapping one; class 2 consists of those pairs that start ($\mu$) in an overlapping configuration and end ($\nu$) in a noninteracting one. The subscript `n$\leftrightarrow$o' stands for `noninteracting$\leftrightarrow$overlapping'. Link-failure factors internal to the pseudocluster (the second product on the right-hand side of Equation~(\ref{accept})) must be accounted for explicitly.

The third product on the right-hand side of Equation~(\ref{accept}) runs over one particular realization ${\cal R}$ of links. It can be thought of as a means of ensuring that the probability of a given particle `pushing' a cluster is the same as its probability of `pulling' the same cluster (or for rotations, ensuring an equal likelihood of effecting a rigid-body rotation in clockwise and anticlockwise directions), and is required to enforce balance. This factor approaches unity for strongly-bound particles, indicating that in this regime multi-particle displacements occur with high probability. The factor becomes small for particles attracting weakly.

Returning to our earlier example, Figure~\ref{figseven}, we see that the only possible contribution to the first factor (of the second argument) of the acceptance criterion (\ref{accept}) can come from bond $jm$ (because in state $\mu$ this bond is `noninteracting' according to our criterion). However, we see that in the proposed new state particles $j$ and $m$ do not overlap, and so their pair energy does not enter this first factor. The latter is therefore unity. Considering the final two factors of  Equation~(\ref{accept}), we see that if the trimer $ijk$ possesses a large binding energy the proposed collective displacement is likely to be accepted. In this case we accept state $\nu$ according to our choice of its diffusion properties.

One strength of the virtual-move scheme is that it allows one to identify groups of particles that move in concert, and to suppress their displacement in order to approximate hydrodynamic damping. A body moving in an overdamped fashion through a fluid possesses translational and rotational diffusion properties that depend on its size and shape. These properties cannot be fully captured by considering only individual or pairwise interactions between the cluster's constituent particles. Simulating the solvent flow that mediates such many-body forces is extremely expensive computationally, and so collective hydrodynamic effects are often neglected when integrating Brownian equations of motion.

Stokes' law states that a cluster of effective hydrodynamic radius $R$ (a measure of the greatest extent of the cluster perpendicular to the direction of motion) possesses a translational diffusion constant $ D_{\rm t}(R) = \Gamma /(3 R)$, and a rotational diffusion constant $ D_{\rm r}(R) = \Gamma /(8 R^3)$, where $\Gamma \equiv \kt (\pi \eta_w)^{-1}$; $\eta_w \approx 10^{-3}$ Pa s is the viscosity of water. We can respect this damping within the virtual-move algorithm by calculating for the chosen pseudocluster its effective hydrodynamic radius $R$, where
\beq
\label{reff}
\left(R-R_0\right)^2= \langle | \left(\mathbf{r}_i - \mathbf{r}_{\rm c} \right) \times \hat{\mathbf{n}} |^2 \rangle.
\eeq
$R_0$ is the monomer radius. The average $\langle \cdot \rangle \equiv n_{\rm {\cal C}}^{-1} \sum_{i=1}^{n_{\rm {\cal C}}}$ runs over all of the $n_{\rm {\cal C}}$ particles $i$ (having coordinates $\mathbf{r}_i$) comprising the pseudocluster (in a particular configuration). The vector $\hat{\mathbf{n}}$ is either the direction of translation or the axis of rotation, as appropriate; the vector $\mathbf{r}_{\rm c}$ is for rotations the position of the centre of rotation, and for translations the centre of mass $ \langle \mathbf{r}_i \rangle$ of the diffusing pseudocluster. This size- and shape-dependent drag becomes increasingly important when the system in question is composed of very polydisperse or anisotropic aggregates.

We enforce this damping by suppressing cluster displacements by a factor $\cal{D}_{\gamma}(R)$, where $\gamma \in \{{\rm t }, {\rm r} \}$ for a translation or rotation as required. We set $\cal{D}_{\rm t}(R)=R_0/R$, and $\cal{D}_{\rm r}(R)=(R_0/R)^3$. 

We consider in Section~\ref{section4} a system of particles with purely attractive interactions and hard-core repulsions, in which case the second class of particle pairs in the set $\langle i j \rangle_{{\rm n}\leftrightarrow {\rm o}}$ does not exist: particles may not start in an overlapping position. In this case Equation~(\ref{accept}) reduces to
\bea
\label{accept2}
 \tilde{W}_{\rm acc}\left(\mu \to \nu|{\cal R}\right) &=&\Theta \left(n_{\rm c} -n_{{\cal C}} \right)\cal{D}_{\gamma}(R) \nonumber \\
&\times& \min \left\{1, \prod_{\langle i j \rangle_{{\rm n} \rightarrow {\rm o}}} e^{-\beta \epsilon_{ij}^{(\nu)}} \prod^{{\cal R}}_{\langle i j \rangle_{{\rm f}}} \frac{ q_{ij}( \nu \to \mu)}{q_{ij}( \mu \to \nu)} \right. \nonumber \\
 &\times& \left.  \prod^{{\cal R}}_{\langle i j \rangle_{\ell}} \frac{ p_{ij}( \nu \to \mu)}{p_{ij}( \mu \to \nu)} \right\}.
\eea
Here the product $\prod_{\langle ij \rangle_{{\rm n} \rightarrow {\rm o}}}$ runs over pairs that do not interact in state $\mu$ (so that $\epsilon_{ij}^{(\mu)}=0$) but possess positive energy (overlap) in state $\nu$.  Barring such overlaps this factor is unity; if such overlaps occur then this factor is zero, and we reject the move. For potentials permitting `soft' overlaps, Equation~(\ref{accept}) does not reduce to Equation~(\ref{accept2}), and overlaps are rejected probabilistically.

The function $\Theta$ is used in conjunction with the early rejection scheme (see Equation~(\ref{virtual_link})). This scheme effects a suppression of the {\em generation} rate of moves of clusters of size $n_{\rm  c}$ by a factor of $1/n_{\cal C}$, ensuring that all particles move with approximately equal frequency.

In Section~\ref{section3} we shall evolve a system of Lennard-Jones discs using both VMMC and a simple Brownian dynamics protocol that neglects collective hydrodynamic effects. The latter enforces a translational diffusion constant for a cluster of size $n_{{\cal C}}$ that scales as $n_{{\cal C}}^{-1}$, and a rotational diffusion constant that scales as the reciprocal of the cluster's moment of intertia. Because disc-disc interactions are isotropic, simply rotating a seed particle about an axis through its centre cannot induce a collective rotation. We therefore use as our basic move a combination of a translation and a rotation. To ensure that the resulting collective translational and rotational diffusion behaves as it would under Brownian dynamics, we generalize the procedure of this section to one that permits the use of distinct {\em real} and {\em virtual} moves. 

This completes our discussion of the the key result of this paper, the virtual-move Monte Carlo scheme. By forming links according to individual bond energies before {\em and after} a proposed move, we displace particles collectively according to individual bond energy gradients without calculating forces explicitly. We ensure that particle positions are updated with approximately equal frequency, and we damp the movement of multi-particle clusters in order to respect Stokes' law. By doing so, we can restore to the Monte Carlo procedure the collective diffusive motion suppressed by making sequential moves of particles in the face of strong, short-ranged interactions. In the following section we test this algorithm against Brownian dynamics simulations. In Section~\ref{section4} we apply VMMC to a model of biological self-assembly.
                                                            
\section{Avoiding unphysical kinetic traps in the face of strong interactions}
\label{section3}

{\em Application to a schematic model of aggregation.} Before applying the virtual-move scheme to an example of self-assembly, we first verify that it evolves a system of particles according to an approximation of natural dynamics. We consider a two-dimensional system of 324 discs of radius $\sigma$. Pairs of discs whose centres are separated by a distance $r$ interact via a Lennard-Jones potential modified to effect a range of attraction that is short compared to a particle's size:
 \beq
 \label{att}
 u(r) =  \epsilon_{\rm b} \Theta \left( r_{\rm c}-r\right)\left[ \cal{L}(\hat{r}/\sigma)- \cal{L}(\hat{r}_{\rm c}/\sigma) \right].
 \eeq
Here $r_{\rm c} \equiv 2.5 \sigma$ and $ \epsilon_{\rm b}= 50 \, \kt$ are respectively the range and strength of the interaction, and $\hat{r} \equiv r -\sigma$ denotes a shifted distance. We have introduced a `Lennard-Jones' function $\cal{L}(x) \equiv 4(x^{-12}-x^{-6})$. The second term in \eq{att} `shifts' the potential to zero at a cutoff distance of $r_{\rm c}$; as a conseqence, the potential minimum is approximately $-35 \, \kt$ (instead of $-50 \, \kt$). Particles occupy about 10\% of the box area.

We shall evolve this system according to a simple Brownian dynamics protocol in which individual particles experience a random force but not a random torque. Collective hydrodynamic effects are ignored. We update particle positions according to the set of equations 
\beq
\gamma \frac{ d{\bf r}_i}{dt} = {\bm \eta}_i + {\bf F}_i,
\eeq
where $i$ labels particles, the forces ${\bf F}_i$ are derived from the potential (\ref{att}), and $\gamma$ is a friction coefficient. The term ${\bm \eta}_i$ is a Gaussian random force with correlations
\beq
\left\langle {\bm \eta}_i(t){\bm \eta}_j(t')\right\rangle = 2 \kt \gamma \, \delta(t-t') \delta_{ij} {\bf 1}.
\eeq
 Such a dynamics effects local motion according to potential energy gradients. It also promotes a buffeting-driven rigid-body translational diffusion of clusters of size $n_{\cal C}$ scaling as $n_{\cal C}^{-1}$, together with a rigid-body cluster rotation scaling as $I_{\cal C}^{-1}$, the reciprocal of the cluster's moment of inertia about the rotation axis (relative to that of a monomer). Note that these results differ from those implied by Stokes' law.

Our aim in this section is to use the virtual move scheme to mimic this dynamics. Because pair interactions are isotropic, collective rotations may not be initiated by rotating a seed particle about an axis through its centre. We therefore use as our basic move a combination of a translation and a rotation. To ensure that the emergent collective motion scales in the appropriate manner it is necessary to generalize the virtual move scheme described in the previous section to one in which we use distinct {\em virtual} and {\em real} moves. We shall use a {\em virtual} move consisting of the superposition of a translation and a rotation to select from our system a pseudocluster. We shall then make a {\em real} move of the pseudocluster which consists of a translation and rotation in general different from those of the virtual move.
\begin{figure}
\includegraphics[width=8cm]{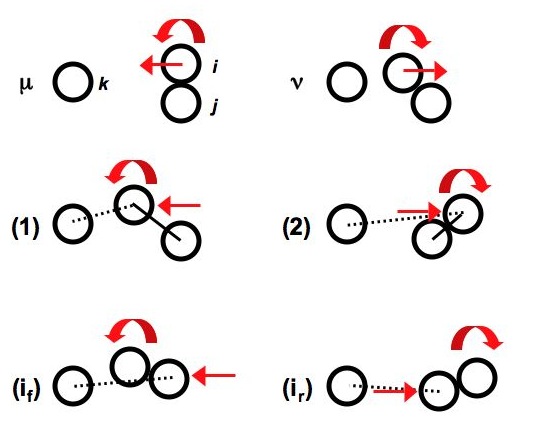} 
\caption{\label{figv2} Generalizing VMMC to allow distinct real and virtual moves. From starting state $\mathbf{\mu}$ the seed particle $i$ is assigned a virtual move defining a leftward translation and a rotation about the ${\bf \hat{z}}$ axis through its centre (here particles are confined to the $x-y$ plane). In this example the link $ij$ forms but $ik$ and $jk$ do not; we therefore record the link-forming probability $p_{ij}(\mu \to {\bf i}_{\rm f})$ and the link-failure probabilities $q_{(i,j)k}(\mu \to {\bf i}_{\rm f})$ [frames $\mathbf{(1)}$ and ${\bf i}_{\rm f}$]. Frame ${\bf i}_{\rm f}$ (see text) shows the intermediate (virtual) state for the forward move. We then return the chosen pseudocluster to its original position (not shown) and execute our chosen real move. Our real move consists of the translational and rotational components of the virtual move with the rotation diminished by the square root of the cluster's moment of inertia about the relevant axis. Here this move yields proposed final state $\nu$. From $\nu$ we record the probability of making link $ij$ by executing the reverse {\em virtual} move, $p_{ij}(\nu \to {\bf i}_{\rm r})$, and the probabilities of failing to form links $ik$ and $jk$ [frames $\mathbf{(2)}$ and ${\bf i}_{\rm r}$]. Frame ${\bf i}_{\rm r}$ shows the intermediate (virtual) state for the reverse move. The acceptance probability for this procedure is given by \eq{virt}.}
\end{figure}

The acceptance rate for such a procedure is correspondingly different from \eq{accept}. To explain its form, we use as an illustration Figure~\ref{figv2}. Here we consider particles $i$, $j$, and $k$, all of which interact. Distances are exaggerated for clarity. From starting state $\mathbf{\mu}$ the seed particle $i$ is assigned a virtual move map consisting of the superposition of a leftward translation and a rotation about the ${\bf \hat{z}}$ axis through the seed's centre (in our two-dimensional example we consider particles to move in the $x-y$ plane). Virtual translation magnitudes and rotation angles are drawn from uniform distributions with respective maxima $\Delta = 0.11 \sigma$ and $\Delta_{\rm rot} \approx 10^{\circ}$. We propose links to all particles with which the seed interacts in state $\mu$. In this example the link $ij$ forms but $ik$ and $jk$ do not; we therefore record the link-forming probability $p_{ij}(\mu \to {\bf i}_{\rm f})$ and the link-failure probabilities $q_{ik}(\mu \to {\bf i}_{\rm f})$ and $q_{jk}(\mu \to {\bf i}_{\rm f})$ [frames $\mathbf{(1)}$ and ${\bf i}_{\rm f}$]. We refer to the state formed by application of the forward virtual move, starting from state $\mu$, as the intermediate state for the forward move, ${\bf i}_{\rm f}$. With the linking scheme finished, we return the pseudocluster $ij$ to its original position (not shown) and execute our chosen real move. 

Our real move also consists of a superposition of a translation and a rotation, but the latter is performed about the ${\bf \hat{z}}$ axis through the centre of mass of the pseudocluster. This allows precise control of rotational and translational degrees of freedom. The direction and magnitude of the translation are identical to that of the virtual translation. We shall account for the diminished diffusion constant of the cluster via the early-rejection scheme, discussed below. We obtain the rotation angle by scaling the magnitude of the virtual rotation by a factor $(I_{\cal C}/n_{\cal C})^{-1/2}$. It is also possible at this stage to scale cluster translations and rotations in order to approximate a hydrodynamic damping. Here we instead employ a scaling designed to mimic Brownian dynamics.

In our example, Figure~\ref{figv2}, the real move produces proposed final state $\nu$. We then ensure reversibility by executing a reverse {\em virtual} move from state $\nu$, and recording the link-forming probability $p_{ij}(\nu \to {\bf i}_{\rm r})$ and the link-failure probabilities $q_{ik}(\nu \to {\bf i}_{\rm r})$ and $q_{jk}(\nu \to {\bf i}_{\rm r})$. The reverse virtual move consists of the forward virtual move with sign inverted (translations occur in the opposite direction; rotations are performed with opposite sense to the forward rotation, but about the same axis relative to the seed particle position). Frame ${\bf i}_{\rm r}$ shows the intermediate (virtual) state for the reverse move.  These factors of $p$ and $q$ ensure that the likelihood of passing from $\mu$ to $\nu$ and back again, via intermediate state ${\bf i}_{\rm f}$ in the forward $(\mu \to \nu)$ direction and intermediate state ${\bf i}_{\rm r}$ in the reverse direction $(\nu \to \mu)$ is such that balance is preserved. The procedure of balancing forward and reverse rates for a particular realization of pseudocluster links {\em and} for specified intermediate virtual states could be termed `superduperdetailed balance'. 

For the case of distinct real and virtual moves we have instead of \eq{accept}
\bea
\label{virt}
&\,&W_{\rm acc}(\mu \to \nu|\cal{R}) = \nonumber \\
&\,& \Theta \left(n_{\rm c} -n_{{\cal C}} \right) \min \left\{1, \prod_{\langle i j \rangle} e^{-\beta\left( \epsilon_{ij}^{(\nu)} -\epsilon_{ij}^{(\mu)} \right)} \right. \nonumber \\
&\times& \left.  \frac{\prod_{\nu \to  {\bf i}_{\rm r}} q_{ij}(\nu \to  {\bf i}_{\rm r}) }{\prod_{\mu \to {\bf i}_{\rm f}} q_{ij}(\mu \to {\bf i}_{\rm f})}
 \prod^{{\cal R}}_{\langle i j \rangle_{\ell}} \frac{ p_{ij}( \nu \to {\bf i}_{\rm r})}{p_{ij}( \mu \to {\bf i}_{\rm f})} \right\}.
\eea
The cutoff $n_{\rm c}$ is drawn from the distribution $\tilde{Q}(n_{\rm c}) \propto n_{\rm c}^{-2}$. One factor of $n_{\rm c}^{-1}$ ensures that particles move with roughly equal frequency. The second factor accounts for the fundamental diffusion rate of $n_{\rm c}^{-1}$ for a cluster of size $n_{\rm c}^{-1}$. This procedure represents a coarse-graining of the dynamics (a cluster of size $n_{\cal C}$ moves one distance unit every $n_{\cal C}$ sweeps, instead of $n_{\cal C}^{-1/2}$ distance units per sweep), and offsets to a considerable degree the waste associated with building large pseudoclusters only to suppress their moves by a large factor. The Boltzmann bond weights include all interactions between the pseudocluster and its environment. The factors of $q$ and $p$ account for links unformed and formed, respectively, during both forward and reverse virtual constructions. Unformed links include those internal to the pseudocluster. The choices of virtual and real moves determine the initial and proposed final states, $\mu$ and $\nu$, and the intermediate states in either direction. 

This scheme allows us to perform cluster rotations and translations whose relative and absolute rates approximate those induced by our Brownian dynamics protocol. While seemingly complicated, the acceptance rate (\ref{virt}) is straightforwardly evaluated during the virtual-move procedure. Furthermore, we find that for late-stage coarsening the acceptance rate is reasonable: for isolated whole-cluster motion the Boltzmann bond weights and the products over the factors $q$ reduce to unity; the link factors $p$ are frequently close to unity, give the strong, short-ranged nature of the interaction (see Appendix B). Moves that result in the motion of single particles are unaffected by the rescaling.
\begin{figure*}[ht]
\includegraphics[width=12cm]{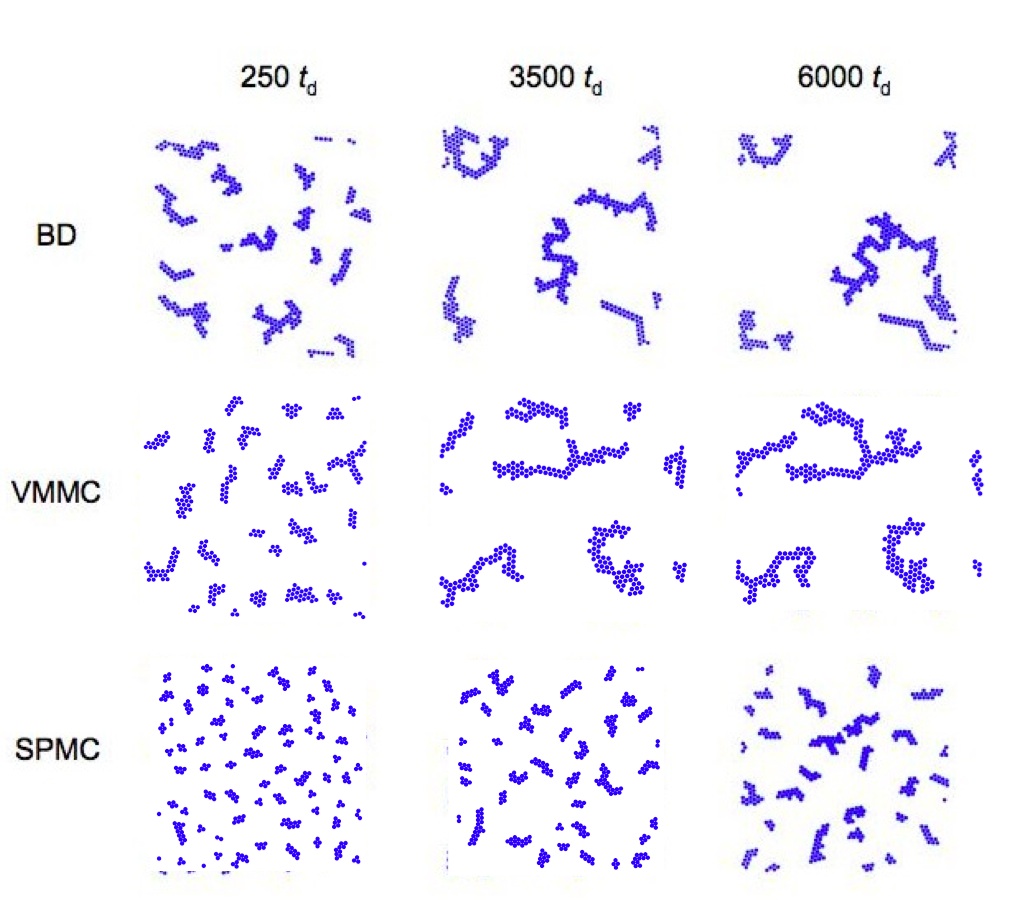} 
\caption{\label{figtraj}  Configurations as a function of time for very attractive discs evolved according to Brownian dynamics (BD), virtual-move Monte Carlo (VMMC) and single-particle Monte Carlo (SPMC) protocols. Cluster sizes and shapes as a function of time are similar under BD and VMMC algorithms. VMMC and SPMC simulations were performed using the same distribution of basic displacements. The acceptance rate for sequential moves of single particles in this regime is low enough to suppress, unphysically, collective modes of relaxation. The computational efficiency of VMMC exceeds that of the BD protocol by more than an order of magnitude.}
\end{figure*}

Starting from a high-temperature configuration we evolve the Lennard-Jones system according to either Brownian dynamics (BD), virtual-move Monte Carlo (VMMC) or single-particle Monte Carlo (SPMC) protocols. We show in Figure~\ref{figtraj} configurations as a function of time obtained under all three protocols. We compare Brownian dynamics and Monte Carlo timescales by reporting times in units of $t_{\rm d}$, the characteristic time for a free monomer to diffuse a length equal to its diameter. We observe similar behaviour as a function of time for BD and VMMC protocols, with the strong, short-ranged interaction inducing clustering. Clusters diffuse and merge, leading to non-compact, kinetically frustrated structures~\cite{footnote}. However, single-particle moves (using the same displacement scale) suppress unphysically the diffusion properties of clusters, leading to much smaller structures on equivalent timescales.

For the system considered here the computational effort required by VMMC is less than that required by the Brownian dynamics protocol. For the latter, a small integration time step is required to maintain numerical stability in the face of the strong, short-ranged interaction; consequently, about 1.2 $\times 10^{5}$ integration steps are required to advance the system one $t_{\rm d}$ unit. By contrast, within the VMMC framework we can employ much larger basic steps, limited only by the requirement that particles take many steps to traverse typical inter-particle distances~\cite{stable}. For our chosen distribution of displacements (whose magnitudes are drawn uniformly from the interval $[0,0.11 \sigma]$) we require only $\approx 1000$ sweeps to advance the system one $t_{\rm d}$ unit. For the trajectories shown, the processor times required to advance the system $(250,3500,6000)$  $ t_{\rm d}$ units are $\approx (3.5,84.5,152.0)$ hours for BD, and $\approx (0.5,6.0,10.0)$ hours for VMMC (processor times reported to the nearest half-hour).

We estimate that our chosen VMMC displacement magnitude is as large as one can employ for this system and density. The clusters formed under VMMC were (for times $t \leq 100 \, t_{\rm d}$) slightly more `ragged' than those formed under BD. Thereafter, we could not distinguish clusters formed by the two protocols (cluster morphologies differ considerably between different trajectories under the same algorithm). Thus, while our implementation of VMMC is not a perfect reproduction of BD, it represents a good approximation thereof. Our calculations (Appendix B) indicate that SPMC would retain a reasonable acceptance rate (and therefore begin to approximate Brownian dynamics) only when the displacement maximum is reduced to less than $\Delta=\sigma/100$. This reduction would diminish the timescale corresponding to a single Monte Carlo sweep by a factor of at least $\sim 100$, and would lengthen simulation times by the same factor.

The considerations of Appendix B indicate that the degree to which VMMC can effect collective motion at a controllable rate depends upon basic displacement scale (for both real and virtual moves), and must be assessed carefully for the model under study. We speculate that the utility of VMMC is greatest for models with short-ranged and anisotropic interactions, for models whose simulation requires a large basic step size (e.g. models defined on a lattice), and for systems whose components are present at low concentration. 

{\em Application to hard particles.} The ability to use distinct virtual and real moves can be used to make collective moves of hard particles without attractive interactions. A suitable algorithm is as follows. We dictate that forward and reverse moves are initiated by the {\em same} virtual move (translation or translation plus rotation) of a given seed particle. The real move consists of either the virtual move (with probability $1/2$) or the virtual move with sign inverted (with probability $1/2$). In the latter case the real translation occurs in the direction opposite the virtual translation, and the real rotation possesses the opposite sense of the virtual rotation. In this manner a given particle can both `pull' and `push' its host cluster, regardless of the strength of energetic interactions. As an example, the three-member cluster shown in Figure~\ref{figreverse} may be moved both rightwards and leftwards via the virtual move depicted in frame {\bf (f)}. This scheme allows one to execute `avalanche' moves of hard particles without regard for the geometry of the avalanche~\cite{krauth}.

\section{Application of VMMC to self-assembly}
\label{section4}
Self-assembly is the process whereby interacting components organize spontaneously into thermodynamically stable patterns or aggregates.  It is the means of formation for many biological structures, including the protein packaging of viruses~\cite{Hagan,Rapaport} to the lipid membranes enclosing cells. The self-assembly of nanometer-scale objects constitutes an important branch of nanotechnology.

In studying self-assembly, it is important to identify the nature of the inter-component interactions that can lead to stable structures with the required symmetry. However, a full understanding of how assembly is effected requires a characterization of its dynamics. Binding that depends on the precise alignment of neighbouring components might ensure thermodynamic stability, but the time required for two bodies to collide in this manner could be prohibitively long. Similarly, very strong contacts contribute to the stability of equilibrium structures but could also transiently stabilize malformed aggregates, thereby impeding equilibration.

Brownian dynamics is perhaps the most natural kinetic model for several examples of biological self-assembly, given the relatively large sizes and small diffusivities of many proteins. However, individual components diffuse much more rapidly than do large-scale structures. A faithful accounting of the faster of these two motions requires, in the face of strong, short-ranged interactions, a small integration time step in order to maintain numerical stability. Long simulation times are therefore required in order to study the assembly of overall structures.

Here we demonstrate that the virtual-move Monte Carlo algorithm can be used to evolve in a computationally efficient manner a collection of self-assembling components possessing strong, short-ranged and angularly specific interactions. The chief advantage of VMMC over Brownian dynamics when applied to such models is that we may use with the Monte Carlo protocol a basic translation step that is not restricted by the width and depth of the pair potential well. With VMMC we face only the less stringent restriction that many steps must be taken to traverse typical inter-particle distances~\cite{stable}. 
\begin{figure}
\includegraphics[width=6cm]{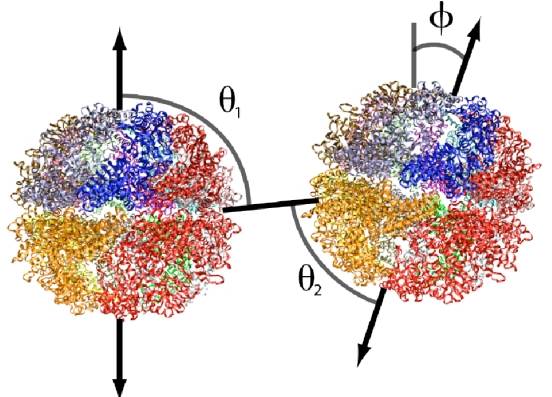} 
\caption{\label{figchap} Geometry for a schematic model of chaperonin self-assembly~\cite{chap1a,chap1b, chap1c}. We use a short-ranged, anisotropic pair potential to mimic the tendency of chaperonins to bind equator-to-equator. We show chaperonin structure determined by homology (image courtesy of Matthew B. Francis and Chad D. Paavola) together with the angles relevant for our chosen interaction. The angle between orientation vectors is $\phi$; the angles between orientation vectors and the inter-unit separation vector are $\theta_1$ and $\theta_2$.}
\end{figure} 

In the presence of ATP and magnesium ions the heat shock protein (Hsp60) from the organism {\em Sulfolobus shibatae} self-assembles in two stages. First, monomers (`sub-units') of the protein assemble into 18-membered, nearly-spherical complexes (`units') of radius 17 nm. Next, units aggregate into extended structures often microns in scale. We refer to the 18-membered complexes as, interchangably, `units' or  `chaperonins'. Experiments with the wild-type protein lead to two distinct types of chaperonin super-structures under subtly different external conditions: two-dimensional sheets with a high degree of hexagonal order, and quasi one-dimensional strings. We focus here on assembly into sheets, the control of which provides a means of engineering organic templates with a high degree of order, potentially useful for electronics devices. Free-floating sheets are also formed by the self-assembly of inorganic nanocrystals~\cite{glotzer_sheets}.

Computer models can help reveal the range of inter-unit attraction strength and specificity required to effect large-scale self-assembly~\cite{patchy1,patchy2}. The detailed interactions between chaperonins are not known. Patterns of polar, nonpolar and charged amino acid side chains exposed by individual chaperonins do not immediately suggest specific regions where two units would strongly bind, nor do experimental results over a range of ionic strength clarify the physical nature of the binding interaction. It is clear, however, that forces between sheet-forming chaperonins favour equatorial contact~\cite{chap2}. We can exploit this information when constructing a simple model of chaperonin self-assembly. 

We build such a model by coarse-graining over the microscopic details of individual units (we consider units to be stable against dissociation into protein monomers). The simplest such approximation is to regard units as spheres without surface detail~\cite{cow}. We mimic the effect of the microscopic unit-unit interactions by endowing spheres with a short-ranged, anisotropic pair potential designed to encourage mutual equatorial contact.

Figure~\ref{figchap} illustrates the geometry of our chosen potential. Each sphere has a polar orientation vector, shown as a double-headed arrow (we assume `up-down' symmetry). We also assume azimuthal symmetry, so that the pair potential does not change if spheres are rotated about their orientation vector (chaperonin units possess 9-fold rotational symmetry about their polar axis). We choose the attraction to be strong if neighbouring orientation vectors are aligned $(\phi=0, \pi)$, and perpendicular to the inter-unit separation vector $(\theta_{1,2}=\pi/2, 3 \pi/2)$. Thus we allow only binding between complementary regions (equator-to-equator), and do not allow, for example, equator-to-pole binding. This reflects the near-perfect alignment observed between chaperonins in sheet-like assemblies, and also the notable absence of disordered aggregates~\cite{chap2}.

Analytically, our chosen potential is
\beq
\label{pot}
\epsilon(\mathbf{r})=-J_{\rm eq} \, \Theta\left(R_{\rm max}- \hat{r}\right)\, \hat{\cal{C}}_1 (\phi) \,\cal{C}_0(\theta_1)\, \cal{C}_0(\theta_2). 
\eeq
The step function $\Theta$ ensures that two spheres interact only when their surfaces are separated by a distance less than $R_{\rm max}$, which we shall vary between $R_0/4$ and $R_0/8$, $R_0$ being the chaperonin radius. Here $\hat{r} \equiv r-2R \geq 0$ is the distance between the surfaces of neighbouring chaperonins. For $\hat{r}<0$ we assume a hard-core repulsion. We parameterize the angular interaction via the `cooperativity' function $\cal{C}_{\alpha} \left( \psi\right) \equiv {\rm e}^{- \left( \cos \psi-\alpha  \right)^2/\sigma^2}$, which rewards an angle $\psi$ if its cosine is within some tolerance (specified by a parameter $\sigma$) of the value $\alpha$.  The symmetrized cooperativity function $\hat{\cal{C}}_{\alpha} \left( \psi\right) \equiv \cal{C}_{\alpha} \left( \psi \right) +\cal{C}_{-\alpha} \left( \psi \right)$ rewards values of $\cos \psi$ near $\pm \alpha$. The first angular factor in (\ref{pot}) encourages the alignment of neighbouring units; the second and third angular factors encourage mutual equatorial contact. We vary $\sigma$ between 0.2 and 0.4.

Here we demonstrate that the virtual-move algorithm (incorporating hydrodynamic damping) described in Section~\ref{section2} can identify several distinct mechanisms of sheet assembly. These range from monomer addition to a single growing cluster (a nucleus), to the binding and merging of separate sheet-like structures. The latter mechanism, which turns out to be a plentiful source of kinetic `traps', is strongly suppressed, using the same displacement scale, by conventional single-particle MC methods. We focus here on typical mechanisms of assembly for our model system, and leave a detailed study of the rates of aggregation as a function of attraction parameters, as well as the mechanisms underlying the formation of hybrid sheet- and string-like structures, for elsewhere.
\begin{figure}
\includegraphics[width=9cm]{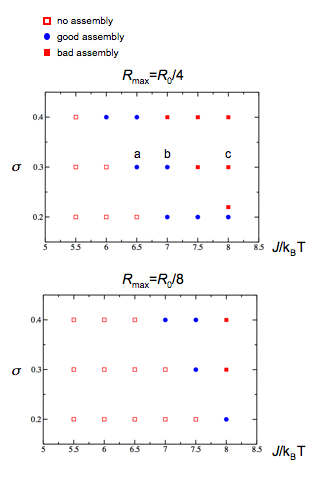} 
\caption{\label{figphase} Kinetic phase diagram for our schematic chaperonin system, evolved using the VMMC algorithm with hydrodynamic damping. We indicate regions of `no assembly' (open squares), `good assembly' (circles) and `bad assembly' (closed squares); see text. A good assembly-bad assembly pair indicates that intermediate building blocks are well-formed, but subsequent multi-particle collisions induce kinetic frustration. Configurations corresponding to parameter sets {\bf a}, {\bf b} and {\bf c} are shown in Figures~\ref{figdynamics1} to~\ref{figdynamics3}.}
\end{figure} 
\begin{figure}
\includegraphics[width=6cm]{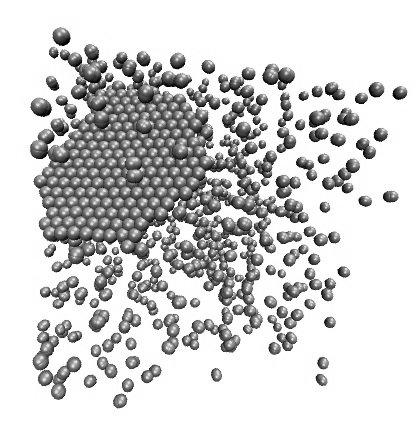} 
\includegraphics[width=6cm]{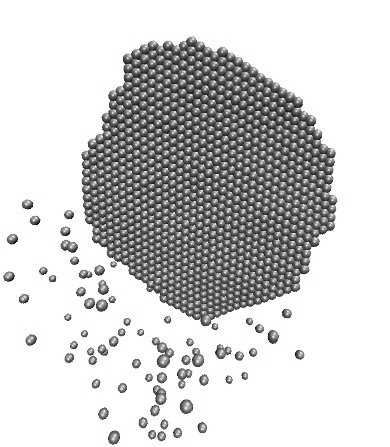} 
\caption{\label{figdynamics1} Configurations obtained after $1.2 \times 10^6$ (top) and $8 \times 10^6$ (bottom) VMMC sweeps for our schematic chaperonin system of spheres with sticky equators of strength $J_{\rm eq} = 6.5 \, \kt$ (parameter set {\bf a} of Figure~\ref{figphase}). At this attraction strength nucleation is sufficiently rare that self-assembly proceeds by the binding of monomers to a single sheet. The resulting structure is relatively well-formed. }
\end{figure} 

By assuming a chaperonin radius of $R_0 = 9$ nm, we obtain from Stokes' law a translational diffusion constant for chaperonin monomers of $D_{\rm t}(R_0) \approx 5 \times 10^{-11}$ m$^{2}$s$^{-1}$ (compare the self-diffusion coefficient of water, $D_{{\rm H}_2 {\rm O}} \approx 5 \times 10^{-9}$ m$^{2}$s$^{-1}$). We draw displacements  from a uniform distribution with magnitude $0.9 R_0$. Our monomer displacement timescale corresponds approximately to $10^{-8}$ s. We fix the relative rates of translation and rotation by imposing Stokes' law for monomers, namely $  3 \, D_{\rm t}(R_0)= 8  R_0^2 \, D_{\rm r}(R_0)$. We ran simulations for $\sim 10^5$ to $\sim 10^7$ MC sweeps, and so probe timescales of the order of seconds. In experiments~\cite{chap1a,chap1b,chap1c}, large-scale assembly is observed on timescales of minutes to hours. Thus we expect our dynamic simulations to detect at least the onset of significant chaperonin self-assembly. 

We start from an initial state consisting of 1000 monomers randomly dispersed and oriented in a three dimensional simulation box with periodic boundaries in each dimension. Units comprised about 0.8 \% by volume of the simulation box, equivalent to a protein concentration of about 5 mg/ml (experimental concentrations of protein range between 1 and 5 mg/ml~\cite{chap2}). We then evolve the system according to the virtual-move scheme with a hydrodynamic damping. Times are quoted in virtual-move Monte Carlo sweeps, with one VMMC sweep corresponding to the (uniform) choice of 1000 seed particles.

In Figure~\ref{figphase} we present a `kinetic phase diagram' of chaperonin assembly, obtained from single trajectories.  We vary interaction strength $J_{\rm eq}$ and inverse angular specificity $\sigma$. For interaction ranges $R_{\rm max} = R_0/4$ (top panel) and $R_{\rm max} = R_0/8$ (bottom panel) we indicate where we find `no assembly' (open squares), `good assembly' (circles) and `bad assembly' (closed squares). We denote by $t_{\rm end}$ the largest time accessed at each parameter point (the `end' of the trajectory). We conclude that `no assembly' has taken place if the largest cluster at time $t_{\rm end}$ possesses fewer than 15 chaperonin units. For those systems for which this is not true, we conclude that assembly is `good' if the constituent monomers of the largest cluster in the system possess on average more than 4.75 `bonds'. We define a particle's bond number as its energy divided by the equatorial coupling, $J_{\rm eq}$. We conclude that assembly is `bad' if, at $t_{\rm end}$, the system's largest cluster possesses more than than 15 members, but fewer than 4.75 bonds per member. 

For those parameter sets exhibiting bad assembly, we identify the largest number of bonds possessed by members of the largest cluster at {\em any} time along the trajectory. If this number is greater than 4.75, we plot a `good assembly-bad assembly' symbol pair. This indicates that while the intermediate building blocks may at some time be well-formed, collisions between these multi-particle structures eventually give rise to an aggregate that is ill-formed. 

A note on timescales: ideally, we would present in a kinetic phase diagram data obtained at fixed Monte Carlo time. Here that is not feasible, because of the broad distribution of relaxation times observed in our chaperonin system. For example, for parameter set $(R_{\rm max}=R_0/4, J_{\rm eq}=8 \, \kb T, \sigma=0.4)$ we observe the aggregation of all particles in $\sim 10^6$ VMMC sweeps. After an equal time, parameter set $(R_{\rm max}=R_0/8, J_{\rm eq}=6.5  \, \kb T, \sigma=0.3)$ exhibits no appreciable assembly (nucleation is very slow). At longer times, however, the latter system is well-assembled. Consequently, we show data at fixed processor times, such that we judged assembly for all parameter sets to be sufficiently far advanced that the ultimate fate of each system is clear. This criterion is clearly arbitrary. However, we believe that the mechanisms of assembly revealed in this manner are qualitatively robust to variations in the criteria used to draw the kinetic phase diagrams.
\begin{figure}
\includegraphics[width=5.5cm]{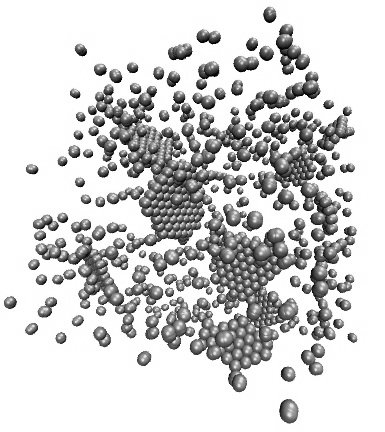} 
\includegraphics[width=5.5cm]{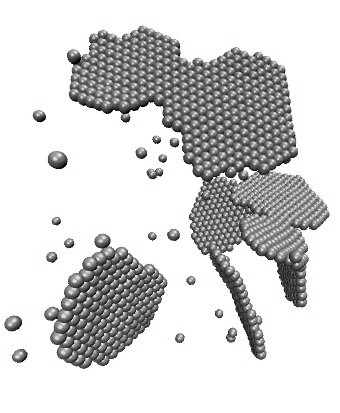} 
\caption{\label{figdynamics2}  Configurations obtained using the VMMC algorithm applied to the chaperonin system with equatorial interaction strength $J_{\rm eq} = 7 \, \kt$ (parameter set {\bf b} of Figure~\ref{figphase}). A modest increase in attraction strength promotes nucleation to a considerable degree. Assembly proceeds both by the addition of monomers to single sheets (top panel, $\sim 0.4 \times 10^6$ sweeps), and via collisions of multi-particle sheets (bottom panel, $\sim 6.3 \times 10^6$ sweeps). Sheets often collide awkwardly, producing ill-formed structures. Here relaxation of these structures is sufficiently rapid that assembly is still `good'. }
\end{figure} 
\begin{figure}
\includegraphics[width=6cm]{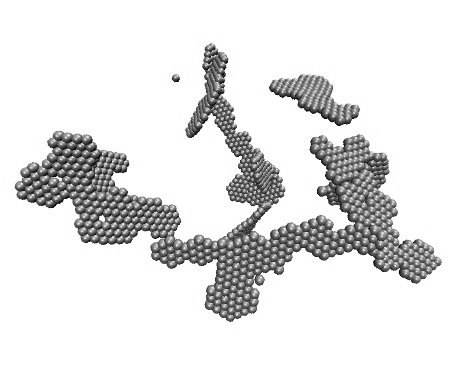} 
\includegraphics[width=6cm]{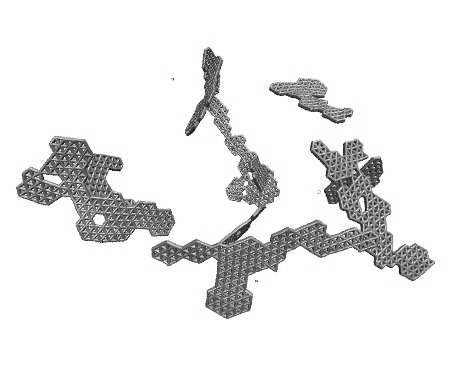} 
\caption{\label{figdynamics3}  Configuration obtained via VMMC for our schematic chaperonin system with equatorial interaction strength $J_{\rm eq} = 8 \, \kt$ (parameter set {\bf c} of Figure~\ref{figphase}). Nucleation is so rapid that multi-particle binding events occur frequently. The resulting structures fail to relax before encountering other such stuctures, leading to aggregates trapped far from equilibrium. Top: excluded-volume view. Bottom: bond view. }
\end{figure} 

The general trend of assembly revealed in Figure~\ref{figphase} indicates that regions of good assembly  occupy a relatively small region of parameter space. This region is defined by a balance between collision rates (controlled largely by density, attraction range and the specificity parameter $\sigma$) and relaxation rate (controlled largely by $J_{\rm eq}$), such that assembled structures form rapidly enough to be observed on the timescales simulated, but not so rapidly that they malform. If this optimal ratio  (`good assembly', circles) is disturbed, we observe over-rapid growth leading to malformed structures (`bad assembly', closed squares), or growth too slow to be observed on the timescales simulated (`no assembly', open squares). For one parameter set we see initially good assembly, where monomers bind into small, well-formed sheets, followed by bad assembly induced by sheets colliding awkwardly and producing malformed structures. The rate of sheet collision is set by Stokes' law, which is respected by the VMMC algorithm. Reducing the attraction range (upper panel to lower panel)  has the effect of reducing collision frequency. We observe an offsetting and a narrowing of the region of good assembly.

It is illuminating to examine the assembly mechanisms observed as we vary only the strength of the equatorial coupling, exemplified by parameter sets {\bf a}, {\bf b} and {\bf c} in the top panel of Figure~\ref{figphase}. In Figure~\ref{figdynamics1} (parameter set {\bf a} of Figure~\ref{figphase}) we show a late-time configuration for units possessing equatorial coupling $J_{\rm eq}=6 \kb T$. Assembly proceeds, following a rare nucleation event, via the binding of monomers to a single sheet-like nucleus. The resulting structure is well-formed and defect-free. The crossover from non-assembly to assembly is rather sharp: at the concentrations used, we observed no assembly within our simulation time at equatorial attraction strength $J_{\rm eq}=5.5 \, \kb T$.
\begin{figure}
\includegraphics[width=5.5cm]{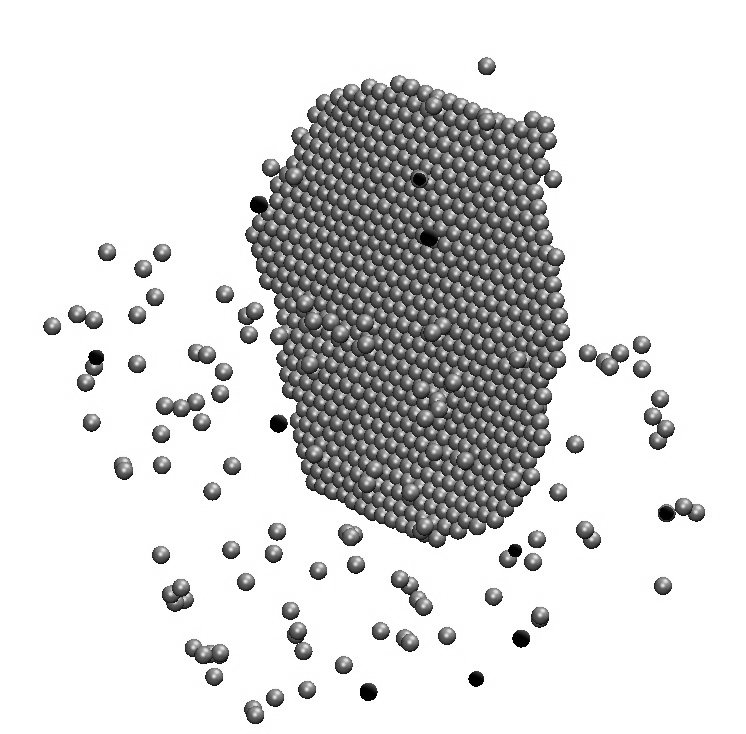} 
\includegraphics[width= 5.5cm]{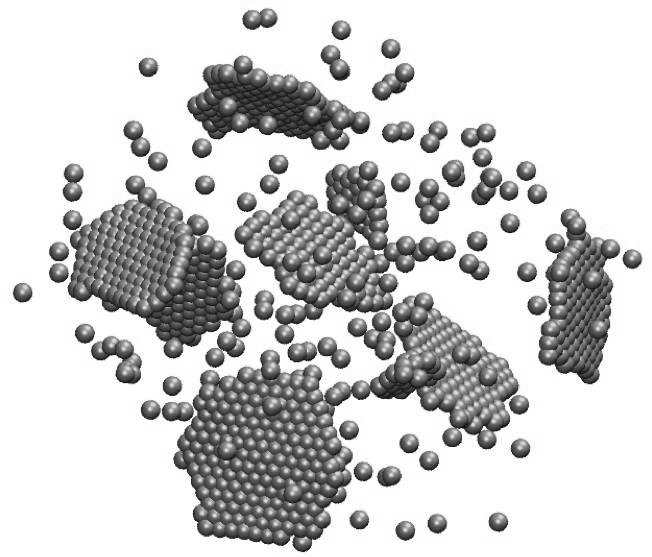} 
\includegraphics[width= 5.5cm]{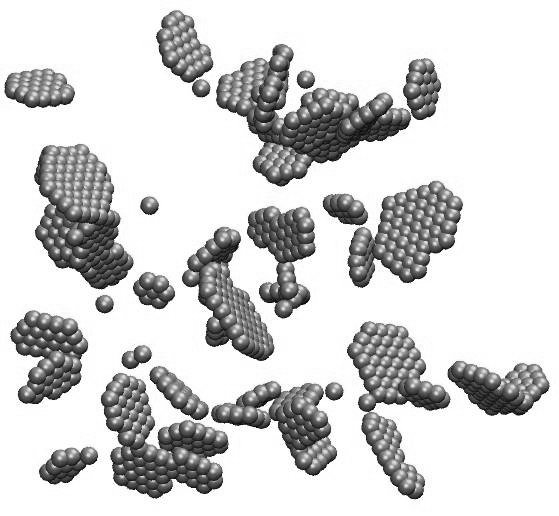} 
\caption{ \label{figspm}  Configurations obtained via a single-particle Monte Carlo algorithm for our schematic chaperonin system with $R_{\rm max} = R_0/4$, $\sigma=0.3$ and (top to bottom) equatorial interactions $J_{\rm eq}=$ 6.5, 7 and 8 $\kb T$. For the displacement distribution employed here (uniform, with maximum displacement 0.9 $R_0$) single-particle moves strongly suppress collective modes of motion, and very few multi-particle collisions take place. For all attraction stengths structures are well-formed, even though VMMC results indicate that for the two larger values of $J_{\rm eq}$ a collective dynamics results in the formation of non-optimal aggregates via multi-particle collisions (see Figures~\ref{figdynamics1} to~\ref{figdynamics3}).}
\end{figure} 

The change in assembly mechanism caused by increasing the unit-unit interaction strength is dramatic.  At a slightly larger interaction strength, $J_{\rm eq} = 7 \, \kt$ (Figure~\ref{figdynamics2}; parameter set {\bf b} of Figure~\ref{figphase}), nucleation proves more rapid. Assembly proceeds via the organization of monomers into multiple sheets. These sheets diffuse according to Stokes' law, and collide with each other. Multi-particle collisions are often awkward, providing an ill-formed template to which monomers bind. In the example shown, however, relaxation is sufficiently rapid that large structures can relax via large-lengthscale fluctuations: assembly is still `good'. 

At still higher attraction strengths, such as $J_{\rm eq} = 8 \, \kt$ (Figure~\ref{figdynamics3}; parameter set {\bf c} of Figure~\ref{figphase}), aggregation is so rapid that nuclei do not have time to relax into low energy sheet-like structures before they bind to other such ill-formed aggregates. In this regime assembly is frustrated kinetically.

These results indicate that the assembly mechanism for our schematic chaperonin model changes considerably with the unit-unit attraction strength. Because these interactions are angularly specific, particles must collide equator-to-equator in order to bind. For insufficiently strong equatorial attractions, random collisions between monomers do not result in stable intermediates, and no assembly is seen. For sufficiently strong attractions, assembly proceeds via the binding of monomers to a single sheet-like nucleus. Aggregates grown in this way are typically well-formed and defect-free. Increasing unit-unit couplings beyond this point slows equilibration. Nucleation is promoted, and collisions between multiple sheet-like nuclei, which occur with a frequency governed by Stokes' law, usually result in awkwardly-bound structures that relax only slowly. There exists a narrow regime of parameter space in which structures formed in this way can relax, via collective fluctuations, into an approximation of a well-assembled sheet. However, at very large attraction strengths nucleation is so rapid that the nuclei themselves are ill formed, leading to disordered aggregates.

For the basic displacement scale considered here, single-particle Monte Carlo techniques suppress unphysically the diffusion of multi-particle structures. When interactions are such that appreciable clustering of particles develops, the dynamics of a system evolved according to single-particle Monte Carlo protocols does not satisfy Stokes' law. As a result, the source of kinetic traps whereby clusters collide and bind awkwardly is strongly suppressed. In Figure~\ref{figspm} we show configurations generated by single-particle translations and rotations applied to systems corresponding to the interaction parameters of Figures~\ref{figdynamics1},~\ref{figdynamics2} and~\ref{figdynamics3} (parameter sets {\bf a}, {\bf b} and {\bf c} of Figure~\ref{figphase}). For all three attraction strengths single-particle moves generate well-formed, isolated clusters. For an attraction strength of $J_{\rm eq} = 6.5$ $\kt$ the VMMC results indicate that the ratio of the rates of cluster growth and diffusion is such that single-particle addition to a single nucleus is the dominant assembly mechanism; single-particle moves naturally capture this dynamics. However, for the larger attraction strengths a dynamical protocol satisfying Stokes' law generates multi-particle collisions, inducing a degree of kinetic frustration that increases with attraction strength. Single-particle moves fail to identify this mechanism. Note that if inter-component interactions are such that assembly must proceed via the interaction of multi-particle structures~\cite{Hagan}, single-particle moves would encounter an unphysical kinetic trap.

\section{Conclusions}
\label{conc}

We have presented a virtual-move Monte Carlo cluster algorithm designed to permit the collective relaxation of particles possessing attractions of arbitrary strength, range and geometry, an important example being self-assembling particles endowed with strong, short-ranged and angularly specific (`patchy') attractions. By calculating pair energies before and after notional (`virtual') moves, we deduce whether particles move individually or in concert. Using an `early rejection' scheme designed to suppress moves of clusters by a factor inversely proportional to the cluster size, we ensure that all particles move with approximately equal frequency. We also ensure that Stokes' law is satisfied for the explicit collective motion of aggregates of arbitrary size and shape.

Our scheme approximates the simultaneous updates of particle positions characteristic of molecular dynamics simulations. Its advantage over the latter lies in the fact that by computing energies before and after virtual moves, one bypasses the need to compute forces and torques explicitly. In addition, one can propose particle displacements that are not limited by the width of the potential well, regardless of the depth of that well. The time savings thus accrued can be considerable (see Section~\ref{section3}), and may be estimated generically as follows. Consider a system of particles evolving over some time $\tau$. We assume that over this period there exist $n_N$ pairwise interactions. The computational effort required by a Brownian dynamics simulation scales as
\beq
\label{cost1}
\cal{C}_{\rm BD} \sim \frac{\tau}{\Delta t} n_N C_{\rm F},
\eeq
where $\Delta t$ is the integration time step and $C_{\rm F}$ is the cost of evaluating forces and torques for each particle pair. The integration time step must be such that a typical particle displacement effects a change in pairwise energy not more than $\kb T$. We therefore have
\beq
\Delta t \sim \frac{1}{2 D}\left(\frac{ \epsilon \sigma}{J} \right)^2,
\eeq
where $D$ is the particle diffusion constant, $\epsilon$ is the width of the potential well in units of the particle radius $\sigma$, and $J$ is the depth of the potential well in units of $\kb T$ (we assume for simplicity a triangular potential well instead of a square well).

The cost of a virtual-move Monte Carlo sweep scales as
\beq
\label{cost2}
\cal{C}_{\rm VMMC} \sim \frac{\tau}{\Delta t_{\rm VMMC}} \alpha n_N C_{\rm E},
\eeq
where $\Delta t_{\rm VMMC}$ is the basic VMMC timescale, $C_{\rm E}$ is the cost of evaluating one pairwise energy, and $\alpha = \cal{O}(1)$ is a parameter accounting for the fact that we may require more than one energy evaluation per particle pair (e.g. when executing a `reverse' virtual move in order to enforce superdetailed balance). The VMMC timescale (derived from the basic displacement scale) is not governed by the width and depth of the potential well, as would be the case for a single-particle Monte Carlo algorithm, but is determined by the less stringent requirement that many steps are taken over typical inter-particle distances. We set the typical displacement to a fraction $f$ of the particle radius. We estimate that
\beq
\Delta t_{\rm VMMC} \sim \frac{\left(p_{\rm acc} f \sigma \right)^2}{2 D},
\eeq
where $p_{\rm acc}$ is the acceptance rate for particle displacements. Comparing Equations~(\ref{cost1}) and~(\ref{cost2}) reveals that
\beq
\frac{\cal{C}_{\rm BD}}{\cal{C}_{\rm VMMC} } \sim \frac{C_{\rm F}}{\alpha C_{\rm E}} \left( \frac{J f p_{\rm acc}}{\epsilon} \right)^2.
\eeq
The savings associated with VMMC become more pronounced the stronger and shorter-ranged is the potential. For the chaperonin model studied here we have $f \sim 1$, $J \sim 10$, $\epsilon \sim 1/5$ and $p_{\rm acc} \sim 0.1$ to 1. The ratio $C_{\rm F} /\left(\alpha C_{\rm E}\right)$ is of order unity ($\alpha \approx 2$, but evaluating forces and torques is more costly than evaluating energies). We obtain therefore $\cal{C}_{\rm BD}/\cal{C}_{\rm VMMC} \sim 10$ to $10^3$. In this estimate we neglect some of the `overhead' of the virtual-move procedure, associated with suppressing the generation rate for moves of large clusters (in order to perform updates of particles with approximately equal frequency). Nonetheless, for attractive interactions of short range this estimate indicates that we might expect considerable time savings using VMMC as opposed to Brownian dynamics.

Further comparisons between the method presented here and Newtonian simulations must be performed before the generic dynamical fidelity of the former can be determined. To this end, a study is underway~\cite{next} in which we compare Brownian dynamics with the VMMC algorithm, where each is used to effect the assembly of idealized protein capsomers into icosahedral virus capsids~\cite{Hagan}.

We expect that the virtual-move algorithm can be used to study the phase behaviour and aggregation mechanisms of a variety of self-assembling systems, both on- and off-lattice.

\begin{acknowledgments}
We thank Edward H. Feng, Matthew B. Francis, Michael Hagan, Robert L. Jack,  Chad D. Paavola, Sander Pronk and Jonathan D. Trent for important discussions. We are grateful to Thomas Ouldridge for identifying an error in the acceptance rate of the algorithm reported in a previous version of this paper [see erratum to J. Chem. Phys. 127, 154101 (2007)]. The similarity in name between our algorithm and the virtual-move parallel tempering scheme of Ref.~\cite{vmpt} is due to our lack of imagination. This work was supported by the US Department of Energy.

\end{acknowledgments}

\section{Appendix A: Using collective moves to facilitate relaxation}
\label{app}
Monte Carlo cluster algorithms~\cite{SW,wu} are used to evolve strongly-attractive particles in order to avoid the suppression of diffusive modes of motion that plague single-particle protocols. Clusters are identified and moved on the basis of properties of particles in the current state of the system, such as energy or degree of proximity. While leading to efficient diffusion of clusters, the internal relaxation of structures evolved in this manner is often under-represented: particles in close proximity or interacting strongly are liable to be moved collectively, and therefore will not re-arrange relative to each other. We demonstrate here that a straightforward modification of these algorithms can be used to efficiently relax, in a collective manner, strongly-interacting particles. However, although equilibration can be facilitated, particle motion does not proceed solely according to local potential energy gradients; the algorithm discussed here should therefore be regarded only as a scheme for sampling equilibrium ensembles. We discuss this point in more detail in Appendix B. The algorithm we describe is similar to that proposed by Troisi, Wong and Ratner~\cite{augmentation}.

We choose as the linking probability 
\beq 
\label{link_prob_1}
p_{ij}(\mu \to \nu) = \textnormal{max}\left[0,1-\exp\left(\bef u_{\rm f}(\e_{ij}^{(\mu)})\right)\right].
\eeq 
Here $\bef$ is a free parameter that functions as a fictitious reciprocal temperature. The term $u_{\rm f}$ is a fictitious potential, and can be chosen for convenience. The simplest choice is to set the fictitious potential equal to the true potential. For some applications a convenient choice is $u_{\rm f}(\e)=\e \, \Theta(r_0-r)$: $ u_{\rm f}(\e)=\e$ within some cutoff distance $r_0$, and zero otherwise. This renders potentials of arbitrary range amenable to the iterative linking scheme described in Section~\ref{section2}. Note that Equation~(\ref{link_prob_1}) depends only on the interaction energy between $i$ and $j$ in the initial state $\mu$.This is a straightforward generalization of the form chosen by Swendsen and Wang~\cite{SW}, and reduces to an off-lattice version of the Swendsen-Wang algorithm in the limit that the fictitious potential and temperature are set equal to their `true' counterparts.

Equation~(\ref{factor3}) gives as the acceptance rate for the move $\mu \to \nu$
\beq
\label{accept0}
W_{\rm acc}(\mu \to \nu)=\textnormal{min} \left(1, \exp\left( -\beta E_{\nu, \mu}+\bef U_{\nu, \mu} \right)\right).
\eeq
Here $ E_{\nu, \mu} \equiv E_{\nu}-E_{\mu}$ and $U_{\nu, \mu} \equiv U_{\nu}-U_{\mu}$, where $E_{\alpha}$ denotes the interfacial energy between pseudocluster $\cal{C}$ and its environment in a given microstate, $ E_{\alpha} \equiv \sum_{{\cal I}_{\alpha}} \e_{ij}$, while $U_{\alpha}\leq 0$ denotes the {\em attractive part} of the {\em fictitious} energy between $\cal{C}$ and its environment, $U_{\alpha} \equiv \sum_{{\cal I}_\alpha} \textnormal{min} \left(0,u_{\rm f}(\e_{ij})\right)$. We have for simplicity set the diffusion term ${\cal D}=1$ and taken the cutoff $n_{\rm c} \to \infty$.

In the case where the inter-particle potential is attractive, $\e_{ij} <0$, and we choose a fictitious potential $u_{\rm f}(\e_{ij})$ equal to the true potential $\e_{ij}$, the procedure we have described is particularly straightforward to implement. Particles are linked according to a simple probability, $p_{ij}=1-e^{-\bef |\e_{ij}|}$, and the ratio of acceptance rates is a simple function of the change in energy resulting from the move. For the move $\mu \to \nu$, the acceptance probability (\ref{accept0}) reduces to
 \beq
\label{accept1}
W_{\rm acc}(\mu \to \nu)=\textnormal{min} \left(1, e^{\left(\bef-\beta\right) \left( E_{\nu}-E_{\mu} \right)}\right).
\eeq 
For infinite fictitious temperature, $\bef =0$, the likelihood of forming links between the seed $i$ and any other particle is zero, and so the algorithm executes single-particle moves with acceptance probability $\textnormal{min} \left( 1, e^{-\beta (E_{\nu}-E_{\mu})}\right)$. For a fictitious temperature equal to the true temperature, $\bef = \beta$, and for attractive interactions, the acceptance probability is unity.

By drawing a fictitious reciprocal temperature in the range $ \bef \in [0,\beta]$ we can interpolate between single-particle moves and rejection-free cluster moves. When applied to a tightly-bound aggregate of particles, the rejection-free scheme does not allow the construction of pseudoclusters smaller than the aggregate. Thus the aggregate cannot relax through moves in concert of its constituents. In other words, the {\em generation} rate for many transitions involving the collective motion of tightly-bound particles is close to zero. By contrast, varying $\bef$ allows one to increase markedly the generation rates of these moves, at the affordable cost of reducing slightly their acceptance rates. As a result, one can choose from the aggregate pseudoclusters of arbitrary size, and thus propose collective internal relaxations in the presence of arbitrarily strong interactions. We refer to this procedure as cluster  `cleaving'. 

\begin{figure}
\includegraphics[width=8cm]{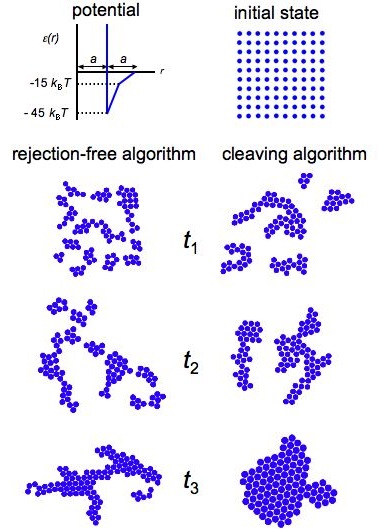} 
\caption{\label{figconfig} 
Configurations from representative trajectories of our test system (see text) at times $t_1<t_2<t_3$. Left panels: single-particle moves plus rejection-free cluster algorithm, leading to a configuration trapped far from equilibrium. Right panels: single-particle moves plus cleaving algorithm. The latter circumvents the kinetic traps associated with rarely testing strong inter-cluster bonds.}
\end{figure}

In Figure~\ref{figconfig} we demonstrate the advantages of the cleaving algorithm over the rejection-free cluster algorithm. We show example configurations from a system 100 hard discs of diameter $a$ in two dimensions, endowed with an attractive piecewise-linear pair potential of range $a$. With the inter-particle separation denoted by $r$, particles experience a hard-core repulsion for $r<a$. The potential (shown in Figure~\ref{figconfig}) increases linearly from its minimum, $\e_0=-45 \, \kt$, at $r=a$ to $-15 \, \kt$ at $r=3a/4$. From $r=3a/4$ the potential increases linearly to zero at $r=2a$. Thereafter, it is zero. This system possesses a thermodynamically stable ground state corresponding to an hexagonal close-packed sheet. 

We find that relaxation is frustrated and that jammed structures persist for the single-particle moves in combination with rejection-free cluster moves. These jammed structures owe their existence to the following mechanism. First, potential energy gradients encourage neighbouring particles to aggregate into dense clumps separated by voids of lower density, a relaxation that can be accomplished readily by single-particle moves. Thereafter, because of the strength of the inter-particle contacts, the rejection-free cluster algorithm fails to propose moves of one clump relative to another. Since the structure does not readily relax in a single-particle fashion -- the steeper gradient of the potential near its minimum disfavours the `jumping' of single particles across a gap -- the system finds itself in a kinetic trap. 

The cleaving algorithm can circumvent this trap. For sufficiently large $T_{\rm f}$ all contacts, regardless of strength, are tested. Here we draw the fictitious reciprocal temperature from a uniform distribution between 0 and $\beta$, $P(\bef) = \beta^{-1} \Theta(\beta-\bef)$, and we use a fictitious potential
\beq
\label{fictpot}
u_{\rm f}(\epsilon) = \left\{ \begin{array}{cc} \epsilon & \qquad (\e \leq \e_0+\delta_{\e})
\\ 0 & \qquad (\e> \e_0+\delta_{\e}) \end{array} \right. .
\eeq
Here $\delta_{\epsilon}=0.1 \, \kt$ is a cutoff energy. The fictitious potential (\ref{fictpot}) returns the true energy $\epsilon$ if $\e$ is within a tolerance $\delta_{\e}$ of the potential minimum, $\e_0$, i.e. if $-45 \, \kt \leq \e \leq -44.5 \, \kt$. If $\epsilon$ is outside this range, the fictitious potential returns zero. This ensures that only those particle that are optimally bound undergo significant collective translations or rotations. The resulting dynamics is unphysical (see Appendix B) but leads to efficient relaxation of the system. Structures formed under the cleaving protocol are typically more compact that those generated by dynamical algorithms.

\section{Appendix B: Monte Carlo dynamics}
\subsection{Single-particle moves}

A Monte Carlo simulation consisting of a sequence of moves of individual particles can approximate natural dynamics for systems in which relaxation is not dominated by diffusive modes of motion. As a simple example, consider a single particle in one dimension in a potential $U(r)$. For sufficiently small basic displacement scale $\Delta$, a Metropolis Monte Carlo trajectory is equivalent to the behaviour described by a diffusive Fokker-Planck equation.

Following Refs.~\cite{mc_dynamics1,mc_dynamics2} we can demonstrate this equivalence for a particle at position $r$ in a potential $U(r)$. The master equation for the particle's motion is
\bea
\label{ma0}
\partial_t P(r;t)&=& \int_{r'} P(r';t) W(r'\to r) \nonumber \\
&-&\int_{r'}  P(r;t) W(r\to r'). 
\eea
Here $\int_{r'} \equiv \int_{r-\Delta}^{r+\Delta} dr' $, $P(r;t)$ is the probability of finding the particle at position $r$ at time $t$, and $W(r \to r')$ is the rate for moving from position $r$ to position $r'$. For the Metropolis acceptance rate the latter reads
\beq
\label{rate0}
W(r \to r')=W_0 \, W_{\rm gen}(\delta r) \min \left(1,{\rm e}^{-\beta \left[U(r')-U(r) \right] } \right).
\eeq
The parameter $W_0$ is a reference frequency, and $W_{\rm gen}(\delta r)=(2 \Delta)^{-1}$ the {\em a priori} probability for choosing from a uniform distribution a step $\delta r \equiv r'-r$ in the range $[-\Delta,\Delta]$. The `$\min$' term encodes the Metropolis acceptance probability.

For sufficiently small $\Delta$ one can expand Equation~(\ref{ma0}) in powers of $\rh \equiv r'-r$. This is most easily done by changing notation from $W(r' \to r)$ to $W(r+\rh;-\rh)$. The latter symbol means the rate for passing from configuration $r+\rh$ $(=r')$ to configuration $r+\rh-\rh=r$, and can be expanded in its first argument: $W(r+\rh;-\rh) =W(r;-\rh)+\rh \partial_r W(r;-\rh)+ \cdots$. Likewise, $P(r',t) = P(r,t) +\rh \partial_r P(r,t) + \cdots$. To second order in $\partial_r$, Equation (\ref{ma0}) reads
\beq
\label{smithers}
\partial_t P(r;t) \approx \partial_r \left( \langle \rh \rangle P(r,t) \right)+\frac{1}{2} \partial_r^2 \left( \langle \rh^2 \rangle P(r,t) \right),
\eeq
with
\beq
\label{moments}
\langle \rh^k \rangle \equiv \int_{-\Delta}^{\Delta} d \rh \,  \rh^k \, W(r;-\rh).
\eeq
The derivative-free term $\int_{r'} P(r,t) W(r;\rh)-\int_{r'} P(r,t) W(r;-\rh)$ vanishes by symmetry. Equation (\ref{smithers}) is a Fokker-Planck equation with drift velocity $v=-\langle \rh \rangle$ and diffusion constant $D=\langle\rh^2\rangle$. Using Equation (\ref{rate0}) we have to first order in $\rh$
\beq
\label{approx_rate}
W(r;-\rh) \approx \frac{W_0}{2 \Delta} \min \left(1,1+\beta \rh \, U'(r) \right).
\eeq
We have assumed that $\rh |U'(r)| \ll 1$. 

The integrals (\ref{moments}) can be evaluated in a piecewise fashion, with the minimum function in Equation (\ref{approx_rate}) returning its first argument when $\rh \geq 0$, and its second argument when $\rh <0$. To second order in $\Delta$ one can calculate that the drift velocity is proportional to the potential gradient, $v =- (\beta/6)   \Delta^2 W_0 \, U'(r) $, and that the diffusion constant is $D = \Delta^2 W_0/3$. This corresponds to a Langevin dynamics satisfying an Einstein (fluctuation-dissipation) relation $-v/D = (\beta/2) \, U'(r) $. Note that normally $v$ is independent of temperature $T$ (not $\propto \beta$ as here), and $D \propto T$ (not independent of $T$). This can be arranged~\cite{mc_dynamics2} by making the fundamental attempt frequency $W_0$ proportional to $T$.

Thus Monte Carlo moves of single particles in a potential take place in a dynamically realistic way, provided the basic step size is such that large changes in energy are not encountered. If large changes in energy are encountered, drift and diffusion cease to be have the forms derived: the first correction to the diffusion term, for example, is $\propto W_0 \Delta^3 U'(r)$. When faced with this problem one must either make the basic step size very small, in which case computational costs can be prohibitive, or recover diffusion by means of explicit collective moves. The algorithms we have introduced address this issue. In the remainder of this Appendix we shall show that the cluster cleaving algorithm (introduced in Appendix A), which forms pseudoclusters by linking particle pairs according to their {\em energies}, corresponds to an unphysical dynamics (a dynamics in which the drift velocity is not simply proportional to the potential gradient), and so should be regarded only as a scheme for sampling equilibrium ensembles. By contrast, the VMMC algorithm, which links particle pairs in a manner consistent with their potential energy {\em gradients}, corresponds to an approximation of realistic dynamics. 

\subsection{Cluster cleaving algorithm}
Here we consider the dynamics of the cleaving algorithm described in Appendix A. We set the fictitious potential $u_{\rm f}$ equal to the true potential $\epsilon$. Let us consider the separation $r\equiv x_j - x_i$ between two otherwise isolated particles $i$ and $j$, which interact via an attractive pair potential $\e_{ij}(r) \leq 0$.

The master equation for the coordinate $r$ is
\bea
\label{ma1}
\partial_t P(r;t)&=& \int_{\hat{r}} P(r';t) W_{\rm c}(r'\to r; \bef) \nonumber \\
&-&\int_{\hat{r}} P(r;t) W_{\rm c}(r\to r'; \bef). 
\eea
Here $\int_{\hat{r}} \equiv \int d \bef \, Q(\bef)  \, \int_{-\Delta}^{\Delta} d \hat{r}$, where $\hat{r} \equiv r'-r$; $P(r;t)$ is the probability of observing a bond separation $r$ at time $t$; and $W_{\rm c}(r \to r'; \bef)$ is the rate at which the cleaving move changes the bond separation from $r$ to $r'$. Recall that $Q(\bef)$ is the distribution from which we draw the fictitious reciprocal potential.

The rate $W_{\rm c}$ (the subscript `c' stands for `cleaving') at which the separation $r$ changes follows straightforwardly from Equations (\ref{link_prob_1}) and (\ref{accept1}). We assume that either particle $i$ or particle $j$ is chosen as a seed, and displaced by the vector $\hat{r}$. Then the bond separation $r$ changes if 1) {\em no} link is proposed between $i$ and $j$, and 2) the Monte Carlo acceptance probability is satisfied. Condition 1) occurs according to Equation (\ref{link_prob_1}) with probability $\min\left[1,\exp\left(\bef \e_{ij}(r)\right)\right]=\exp \left(\bef \e_{ij}(r)\right)$. Criterion 2) is satisfied with a likelihood equal to the right-hand side of Equation (\ref{accept1}). Hence
\bea
W_{\rm c}(r \to r')=\left( 2 \Delta \right)^{-1} W_0 \,  \exp \left[\bef \e_{ij}(r)\right]\nonumber \\
\times \min \left(1,{\rm e}^{-(\beta-\bef) \left[\epsilon_{ij}(r')-\e_{ij}(r) \right] } \right).
\eea
Once again, $W_0$ is a reference frequency, and again we assume that we can expand the master equation (\ref{ma1}) to second order in the small displacement $\Delta$. We obtain the Fokker-Planck equation
\beq
\label{fp2}
\partial_t P(r;t) \approx-\partial_{r} \left(v_{\rm eff} P(r;t)\right) +\frac{1}{2} \partial_{r} \left(D \partial_{r} P(r;t ) \right),
\eeq
with effective drift velocity
\beq
\label{drift1}
v_{\rm eff} = v- \frac{1}{2} \partial_{r} D.
\eeq
The `bare' drift velocity $v$ is
\beq
v = - \frac{\Delta^2}{6} \hat{W} \cdot \left(\beta - \bef \right)  \partial_{r} \e_{ij}(r) ,
\eeq
and the diffusion constant is
\beq
\label{diff1}
D= \frac{\Delta^2}{3} \hat{W}.
\eeq
Here we have defined the position-dependent rate
\beq
\label{awkward}
\hat{W} = W_0 \int d \bef Q(\bef)  \, {\rm e}^{\bef \e_{ij}(r)},
\eeq
which contains an integral over $\bef$ (this integral acts on any $\bef$-dependent factors to its right). By virtue of the position-dependence of (\ref{awkward}), the effective drift velocity is not simply proportional to the negative of the potential gradient. Instead, it is biased by a term depending on the exponential of the local bond energy:
\beq
v_{\rm eff} = - \frac{\Delta^2}{6} W_0 \beta \,  [\partial_{r} \e_{ij}(r) ] \int d \bef Q(\bef)  \, {\rm e}^{\bef \e_{ij}(r)}.
\eeq
This bias is not consistent with physical dynamics. The cleaving algorithm evolves the system according only to a rough approximation of true Langevin dynamics: a particle's drift velocity is not simply proportional to the potential gradient it experiences, nor is the diffusion constant (\ref{diff1}) independent of position. Only in the (undesirable) limit of single-particle moves, $Q(\bef) = \delta(\bef)$, is local physical dynamics restored. Note also that in the convenional rejection-free limit, $Q(\bef) = \delta(\bef-\beta)$, drift and diffusion are both suppressed by a factor ${\rm e}^{\beta \e_{ij}(r)}$, demonstrating that in the limit of strong attractions little motion is possible.

The root of these difficulties is the fact that pseudoclusters are built according to pair energies, and not energy gradients. However, drift and diffusion still satisfy the required Einstein relation for evolution to equilibrium,
\beq
- \frac{v_{\rm eff}}{D} = \frac{ \beta}{2}  \, \partial_{r} \e_{ij}(r).
\eeq
This condition is equivalent to the master equation~(\ref{ma1}) satisfying balance.
\subsection{VMMC algorithm}
Next we turn to VMMC algorithm described in Section~\ref{section2}. With similar notation to before, the master equation for the separation $r$ between two otherwise isolated particles $i$ and $j$ is 
\bea
\label{ma2}
\partial_t P(r;t)&=& \int_{\hat{r}} P(r';t) W_{\rm v}(r'\to r) \nonumber \\
&-&\int_{\hat{r}} P(r;t) W_{\rm v}(r\to r'),
\eea
where $\int_{\hat{r}} \equiv \int_{-\Delta}^{\Delta} d \hat{r}$. The rate $W_{\rm v}$ (the subscript stands for `virtual') at which the separation $r$ changes follows from Equation (\ref{virtual_link}). We assume that the virtual displacement of particles $i$ relative to $j$ results in a change in the bond separation $r$ by $\hat{r}$. This change is accepted if 1) {\em no} link is proposed between $i$ and $j$, and 2) the Monte Carlo acceptance probability is satisfied. Condition 1) occurs according to Equation (\ref{virtual_link}) with probability
\beq
\label{nolink}
q_{ij}(r \to r') =\min \left[1, \exp\left(\beta \e_{ij}(r) - \beta  \e_{ij}(r') \right) \right].
\eeq
If condition 1) is satisfied then condition 2) is automatically true provided that no particle overlaps occur. Thus
\beq
W_{\rm v}(r \to r')=\frac{W_0}{2 \Delta} \, q_{ij}(r \to r').
\eeq
In a regime in which the energy change induced by the basic displacement is small, we may expand Equation (\ref{ma2}) in powers of $\Delta$. This procedure yields a Fokker-Planck equation with a physically realistic drift velocity
\beq
\label{drift2}
v = -\frac{\Delta^2}{6} W_0 \beta \, \partial_{r} \e_{ij}(r) ,
\eeq
and a position-independent diffusion parameter
\beq
\label{diff2}
D = \frac{\Delta^2}{3} W_0.
\eeq
The rates of drift and diffusion of the relative coordinate $r$ per Monte Carlo sweep are twice the values given by Equations~(\ref{drift2}) and~(\ref{diff2}), since both $i$ and $j$ are selected as `seed' on average once per sweep. These results agree with those obtained by assuming that $i$ and $j$ are subject to a Brownian motion described by the equations
\bea
\gamma \dot{x}_i = -\partial_{x_i} \epsilon(r) +\eta_i ;\nonumber \\
\gamma \dot{x}_j = -\partial_{x_j} \epsilon(r) +\eta_j,
\eea
provided that the friction coefficient $\gamma$ is related to the Monte Carlo attempt frequency $W_0$ by $\Delta^2 W_0 \gamma = 6 \kt$. The Gaussian white noise satisfies $\langle \eta_i(t) \eta_j(t') \rangle = 2 k_{\rm B}T \gamma \delta_{ij} \delta(t-t')$.
 
We next consider the motion of the centre of mass $R \equiv \frac{1}{2} (x_i+x_j)$ of the dimer $ij$. The centre of mass position may be changed if 1) no link is formed between $i$ and $j$, in which case one particle moves relative to the other, or 2) if a link is formed between the particles, in which case $i$ and $j$ may move collectively. The master equation for $R$ is consequently
\bea
\label{maR}
\partial_t P(R)&=& \frac{W_0}{ \Delta} \int_{r'} P\left(R-\frac{\hat{r}}{2}\right) q_{ij}(r \to r') \nonumber \\
&-& \frac{W_0}{ \Delta} \int_{r'} P(R) q_{ij}(r \to r') \nonumber \\
&+& \frac{fW_0}{ \Delta} \int_{r'} P\left(R-\frac{\hat{r}}{s}\right) p_{ij}(r \to r') \min \left( 1,\frac{\hat{p}}{p}\right)\nonumber \\
&-& \frac{fW_0}{ \Delta} \int_{r'} P(R) p_{ij}(r \to r')  \min \left( 1,\frac{\hat{p}}{p}\right).
\eea
The first two lines of \eq{maR} arise from relative moves of $i$ and $j$, while the second two lines describe moves of $i$ and $j$ in concert. The frequency parameter $f$ permits an adjustment of the rate of collective moves relative to those of single particles (it is controlled by the early-rejection scheme described in Section~\ref{section2}), and the scale factor $s$ quantifies the chosen scaling of the collective displacement relative to that of a monomer. The `min' function enforces the requirement that forward and reverse collective moves must be initiated by a given seed particle moving in the forward and reverse directions; we have defined $p \equiv   p_{ij}(r \to r + \hat{r})$ and $\hat{p} \equiv p_{ij}(r \to r-\hat{r})$. The drift of the centre of mass therefore satisfies
\bea
\label{drift3}
\langle R^2 \rangle &=& \frac{W_0}{\Delta} \int_{r'} \left( \frac{\hat{r}}{2} \right)^2 (1-p) \nonumber \\
&+&\frac{f W_0}{\Delta} \int_{r'} \left( \frac{\hat{r}}{s} \right)^2 p \cdot \min \left( 1,\frac{\hat{p}}{p}\right).
\eea
\begin{figure}[ht]
\includegraphics[width=8cm]{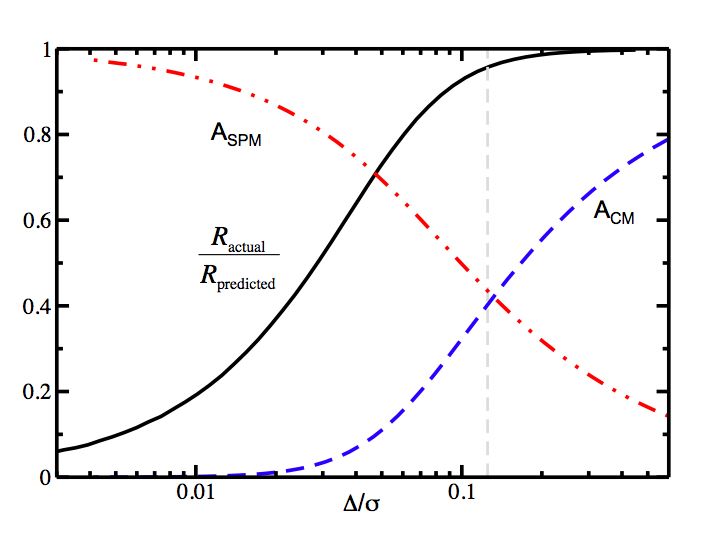} 
\caption{\label{figacc} Acceptance rates (vertical axis) as a function of maximum displacement $\Delta$ (horizontal axis) for two particles interacting with the Lennard-Jones-esque potential discussed in Section~\ref{section3}. Simulations in Section~\ref{section3} were performed using $\Delta = 0.11 \sigma$. The solid line shows the ratio of actual cluster displacements to the displacement predicted on the grounds of cluster-move proposal rate, $R_{\rm actual}/R_{\rm predicted}$. The deviation from unity is due to the superdetailed balance factor. This ratio is close to unity for the displacement scale we use, indicating that little unwanted suppression of collective motion occurs. We show also the fraction of moves for which relative particle displacements are accepted ($A_{\rm SPM}$, dot-dashed line), and the fraction of moves for which the particles are displaced collectively  ($A_{\rm SPM}$, curved dotted line). }
\end{figure}

When $p$ is small (corresponding to a small change in energy upon the proposed move), the collective diffusion of the dimer is dominated by single-particle moves, and hence by the first term in~\eq{drift3}. When $p$ approaches 1 (corresponding to a large change in bond energy upon the proposed move), collective diffusion induced by single-particle moves is strongly suppressed.  Within the VMMC algorithm this suppression of diffusion is in principle compensated by explicit collective moves of $i$ and $j$, described by the second line of~\eq{drift3}, provided that $f$ and $s$ are chosen accordingly. However, the requirement of superdetailed balance suppresses the rate of collective moves by a factor $ \min \left( 1,\hat{p}/p\right)$. If this factor deviates significantly from unity it may by necessary to adjust the frequency $f$ or displacement scale $s$ of collective moves to restore the desired `physical' diffusion properties of the cluster centre of mass. 

For the Lennard-Jones system discussed in Section~\ref{section3} we may determine the unwanted suppression of cluster diffusion, due to the superdetailed balance factor, for two particles placed initially at a separation such that their energy of interaction is as favourable as possible. We pick one particle and propose a displacement of that particle. We draw proposed displacements $\hat{r}$ uniformly from the interval $[0, \Delta]$. We compute the change in pair energy resulting from this proposed displacement, and form a link between particles with the virtual-move probability $p(r \to r+ \hat{r})$. If a link does not form we accept the move of one particle with respect to the other. If a link forms, we reject the proposed displacement. We instead increment by $\hat{r}^2$ the `predicted' squared displacement, $R^2_{\rm predicted}$: this is the displacement of the {\em dimer} expected if every linking event results in collective motion. We then compute the reverse linking weight, $\hat{p}=p(r \to r- \hat{r})$, and with probability $\min(1,\hat{p}/p)$ increment by $\hat{r}^2$ the `actual' squared displacement $R_{\rm actual}^2$. For a given $\Delta$ we perform this procedure $5 \times 10^6$ times. We show in Figure~\ref{figacc} as a function of $\Delta$ the ratio of actual to predicted cluster displacements for moves in which a link forms between the two particles. We show also the fraction of moves in which particles move singly $A_{\rm SPM}$ (no link forms) and collectively $A_{\rm CM}$ (a link forms and the ratio $\hat{p}/p$ permits movement of the dimer). Simulations in Section~\ref{section3} were performed using a basic translational displacement scale of $\Delta = 0.11 \sigma$. In this regime the requirement of balance causes little unwanted suppression of collective motion. It should be noted that $\hat{p}/p$ is in general very close to unity for subsequent links made within large clusters: because virtual moves contains a rotational component, particles recruited iteratively to the pseudocluster tend to move larger distances (in both forward and reverse directions) as the iteration progresses.

The upper limit of $\Delta$ is governed by the requirement that motion be properly `diffusive' on length scales set by typical particle separations. It is likely that the utility of VMMC is greatest when typical particle separations are large compared with particle diameters, which in turn are large compared with potential well widths. This is the case for models of self-assembling proteins, whose real-life counterparts are typically present at low concentrations. It should also be noted that the effective collective motion induced by moves of single particles produces cluster diffusivities akin to those of Brownian dynamics. If collective diffusion within VMMC is assigned a different scaling (e.g. in order to satisfy Stokes' law), it should be verified that the two mechanisms of cluster diffusion do not `compete'. For a given model, careful consideration should be given to the choice of displacement scales (for both real and virtual moves) and attempted versus actual cluster displacements, in order to ensure that the desired motion of clusters is approximated.

The generalization to many particles of the argument presented in this section is not straightforward. The simple comparison performed in Section~\ref{section3} indicates that for our chosen system VMMC represents a reasonable approximation of Brownian dynamics. The utility of VMMC in other cases must be assessed by testing against established methods, such as theoretical results (e.g. the solutions to Langevin equations, for sufficiently simple models) or Newtonian simulation.


\begin{thebibliography}{99}
\bibitem{footnote0} A simple Brownian dynamics protocol subjects every particle at each time step to a random buffeting force. Assume that each such `kick' changes the position of a particle by on average one unit of distance. Then the typical kick acting on a single particle in a cluster of size $n_{\cal C}$ changes the position of the cluster centre of mass by $n_{\cal C}^{-1}$ distance units. Because kicks are uncorrelated, the total change in cluster position resulting from kicks to each particle (per time step) scales as $\sqrt{n_{\cal C}} \cdot n_{\cal C}^{-1} =n_{\cal C}^{-1/2}$. The translational diffusion constant of the cluster is proportional to the square of this result, $D_{\rm t}^{\rm BD}(n_{\cal C}) \propto n_{\cal C}^{-1}$. A single-particle Monte Carlo simulation results in a translational diffusion constant $D_{\rm t}^{\rm SPMC}(n_{\cal C}) \sim \bar{p}_{\rm acc} \, n_{\cal C}^{-1}$, where $\bar{p}_{\rm acc}$ is the mean acceptance rate for moves of particles in the cluster.
\bibitem{footnote2} Note that for particles on a lattice (whose sites are no smaller than the particles) this procedure is not required to ensure balance: a move in which a seed particle `pushes' a cluster can always be reversed by a different seed particle pushing back in the opposite direction. However, dynamical correctness requires that the superdetailed balance requirement be enforced (for example, we do not want to move, in concert, adjacent particles with no energy of interaction).
\bibitem{footnote} Hydrodynamic forces mediated by the exchange of momentum between solvent particles can induce long-ranged attractions between solute particles. These forces can suppress the aggregation of clusters via contact between large surfaces, effectively enhancing binding at cluster `endpoints'. This effect leads to `stringier' clusters than produced by standard Brownian dynamics schemes~\cite{ard1}, and may be treated within the framework of `stochastic rotation dynamics'~\cite{ard2}. We do not consider such effects here. However, we speculate that the essence of this phenomenon may be captured by scaling the real move according to the pseudocluster's environment.
\end{thebibliography}
\end{document}